\documentclass[aps,twocolumn,superscriptaddress,showpacs]{revtex4-1}
\usepackage{amsmath,amssymb,graphicx}

\usepackage[T1]{fontenc}
\usepackage{chancery}

\usepackage{xcolor}
\usepackage{physics}
\usepackage[bottom]{footmisc}

\IfFileExists{newtxtext.sty}   {\usepackage{newtxtext,newtxmath}}
{\IfFileExists{stix.sty}
	{\usepackage{stix}}
	{\IfFileExists{mathptmx.sty}
		{\usepackage{mathptmx}}{} } }

\usepackage{textcomp}

\usepackage{bm}

\IfFileExists{siunitx.sty}{\usepackage{booktabs,siunitx}}{}

\usepackage{color}
\definecolor{LinkColor}{rgb}{0.256,0.439,0.588}
\usepackage{hyperref}
\hypersetup{
	pdfauthor={Tsung-Cheng Lu},
	colorlinks=true,
	citecolor=LinkColor,
	linkcolor=LinkColor,
	urlcolor=LinkColor
}

\usepackage{pifont}
\usepackage{epstopdf}
\usepackage{caption}
\usepackage{subcaption}
\renewcommand{\raggedright}{\leftskip=0pt \rightskip=0pt plus 0cm}
\newcommand{\beq}{\begin{eqnarray}}
\newcommand{\eeq}{\end{eqnarray}}
\newcommand{\mA}{\mathcal{A}}
\newcommand{\mB}{\mathcal{B}}

\begin{document}
	
	\title{Characterizing Long-Range Entanglement in a Mixed State\\
	Through an Emergent Order on the Entangling Surface}


	

	
	\author{Tsung-Cheng Lu} 
		\affiliation{Perimeter Institute for Theoretical Physics, Waterloo, Ontario N2L 2Y5, Canada}
	\affiliation{Department of Physics, University of California at San
		Diego, La Jolla, CA 92093, USA}

		\author{Sagar Vijay} 
	\affiliation{Department of Physics, University of California, Santa Barbara, CA 93106, USA}
	\begin{abstract}

Topologically-ordered phases of matter at non-zero temperature are conjectured to exhibit universal patterns of long-range entanglement which may be detected by a mixed-state entanglement measure known as entanglement negativity. We show that the entanglement negativity in certain topological orders can be understood through the properties of an emergent symmetry-protected topological (SPT) order which is localized on the entanglement bipartition. This connection leads to an understanding of ($i$) universal contributions to the entanglement negativity which diagnose finite-temperature topological order, and ($ii$) the behavior of the entanglement negativity across certain phase transitions in which thermal fluctuations eventually destroy long-range entanglement across the bipartition surface. Within this correspondence, the universal patterns of entanglement in the finite-temperature topological order are related to the stability of an emergent SPT order against a symmetry-breaking field.  SPT orders protected by higher-form symmetries -- which arise, for example, in the description of the entanglement negativity for $\mathbb{Z}_{2}$ topological order in $d=4$ spatial dimensions -- remain robust even in the presence of a weak symmetry-breaking perturbation, leading to long-range entanglement at non-zero temperature for certain topological orders.

	\end{abstract}

	\maketitle

Strongly-interacting quantum phases of matter at zero temperature can exhibit universal patterns of long-range entanglement \cite{wen1989vacuum,wen1990ground,wen1990topological}, which may be used to store and manipulate quantum information.  
Many of these quantum phases, such as topological orders in two and three spatial dimensions, cannot function as self-correcting quantum memories at non-zero temperature \cite{dennis2002,nussinov2008autocorrelations,nussinov2009sufficient,bravyi2009no_go,hastings2011,yoshida2011,Poulin2013,brown2016review}. It remains of interest to characterize the patterns of long-range entanglement in quantum many-body systems that can survive the presence of thermal fluctuations. 

A thermal density matrix $\rho\sim e^{-\beta H}$ of a quantum many-body system described by a local Hamiltonian $H$ is said to exhibit topological order if it cannot be prepared from a classical mixed state using finite-depth local quantum circuits \cite{hastings2011}. It was recently proposed \cite{Lu_topo_nega_2020} that topological order in a thermal state can be detected by the {entanglement negativity} \cite{peres1996,horodecki1996,eisert99,vidal2002}, a mixed-state entanglement measure which has been studied extensively in quantum many-body systems   \cite{audenaert2002entanglement,Eisler_2015,Tonni_negativity_2015,Eisler_2016_free_lattice,Bianchini_2016_free_boson,Shapourian2017,Shapourian2018, calabrese2012,tonni_quench_cft,negativity_large_c_2014_kulaxizi,calabrese2015,Tonni_negativity_cft_2015, Bose_2009_spin_chains,Calabrese_2013_critical_ising,random_spin_chain_2016,gray2018fast,negativity_ising_chain_2018,negativity_random_singlet, vidal2013,castelnovo2013,wen2016topological,wen2016edge,castelnovo2018,nega_hybrid_fisher,nega_hybrid_shi}; in particular, 
 topologically-ordered mixed states were conjectured \cite{Lu_topo_nega_2020} to possess a universal, constant contribution to the entanglement negativity which quantifies the long-range entanglement that cannot be removed by a finite-depth, local quantum circuit. 
  This correction, termed the \emph{topological entanglement negativity}, coincides with the topological entanglement entropy \cite{levin2006detecting,Kitaev06_1} at zero temperature, though it is known that the latter fails as a diagnostic for topological order in a mixed state. Outstanding questions remain regarding (i) the universality of the topological entanglement negativity within a finite-temperature topological order, and (ii) its behavior across a thermal phase transition in which the topological order is destroyed.  

\begin{figure}
	\centering
	\begin{subfigure}[b]{0.46\textwidth}
		\includegraphics[width=\textwidth]{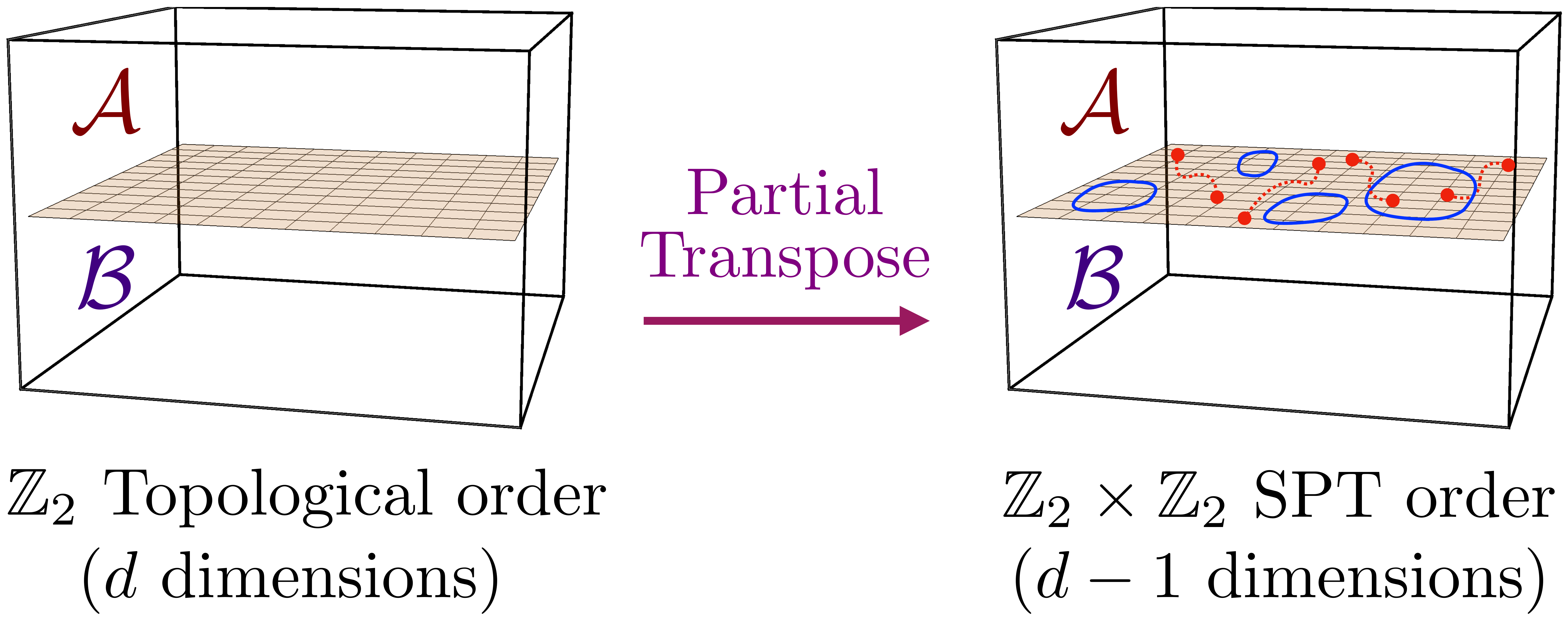}
	\end{subfigure}
	\caption{The thermal density matrix for certain topological orders after partial transposition on a subregion $\mathcal{A}$ -- denoted $\rho^{T_{\mathcal{A}}}$ -- can be related to an emergent symmetry protected topological (SPT) order on the boundary of $\mathcal{A}$.  Eigenvalues of $\rho^{T_{\mathcal{A}}}$ are related to ``strange correlators'' $\bra{ +} O  \ket{\psi}$, where $\ket{+}$ is a symmetric trivial state and $\ket{\psi}$ is an SPT ordered state.  We argue that long-range order in these strange correlators gives rise to a non-zero topological entanglement negativity.  This schematic correspondence for the $\mathbb{Z}_{2}$ topological order in $d$ spatial dimensions is shown. 
	}    \label{fig:spt}
\end{figure}

In this work, we make progress towards answering these questions by demonstrating a connection between the entanglement negativity in certain topological quantum orders, and the properties of an emergent, symmetry-protected topological (SPT) order \cite{spt_1d_2011,spt_2011}  localized on the entanglement bipartition.  Specifically, we show that the entanglement negativity is determined by the ``strange correlator'' \cite{You_strange_correlator_2014} for this emergent SPT order 
(see Fig.\ref{fig:spt}).  We argue that the stability of the SPT phase, as characterized by the presence of long-range order in these strange correlators, is intimately related to the robustness of the topological entanglement negativity in the finite-temperature topological order.   In addition, we show that proliferating topological excitations near the entanglement bipartition can drive a phase transition in which long-range entanglement  across the bipartition is destroyed.  We derive universal scaling forms for the topological entanglement negativity by relating this ``disentangling''  thermal phase transition to a zero-temperature phase transition between the emergent SPT order and a trivial state.  

The connection that we identify between an emergent SPT order and a topologically-ordered mixed state may be understood heuristically as follows. In certain gapped topological orders at zero temperature, the ``partially-transposed” density matrix with respect to a subsystem — an essential operation in the calculation of the entanglement negativity of that region — can be regarded as a wave function in which gapped excitations of the topological order have proliferated near the entanglement bipartition, with relative configurations of the excitations weighted by their statistical braiding phase. In a dual description, this wave function describes an SPT order where the protecting symmetry is inherited from the gauge symmetry of the bulk topological order, when restricted to the entanglement bipartition. When the bulk topological order is at a non-zero temperature, the excitations in this state are no longer localized to the entangling surface, and the partially-transposed density matrix may be related to an emergent SPT state which is acted upon by a symmetry-breaking field. While most SPT orders are immediately destroyed by such a perturbation, if the symmetry to be broken is a one-form symmetry \cite{higher_form_2015}, the SPT order can remain robust  below a finite strength of the symmetry-breaking field, resulting in a stable long-range entanglement below a finite critical temperature $T_c$.  This scenario occurs in the $\mathbb{Z}_{2}$ topological order in $d=4$ spatial dimensions as well as in $d=3$ dimensions with $\mathbb{Z}_{2}$ charge excitations forbidden.  

While we focus on $\mathbb{Z}_{2}$ topological order in various dimensions throughout this work, our characterization of long-range entanglement in a mixed state through the stability of an emergent SPT order also holds for $\mathbb{Z}_{n}$ topological orders as well as certain fracton orders \cite{New_Paper}.  The applicability of this correspondence to other topological orders, as well as the behavior of the negativity across phase transitions in which the entire bulk topological order is destroyed by thermal fluctuations remain important open questions of our work.

 \begin{figure}
	\centering
	\begin{subfigure}[b]{0.38\textwidth}
		\includegraphics[width=\textwidth]{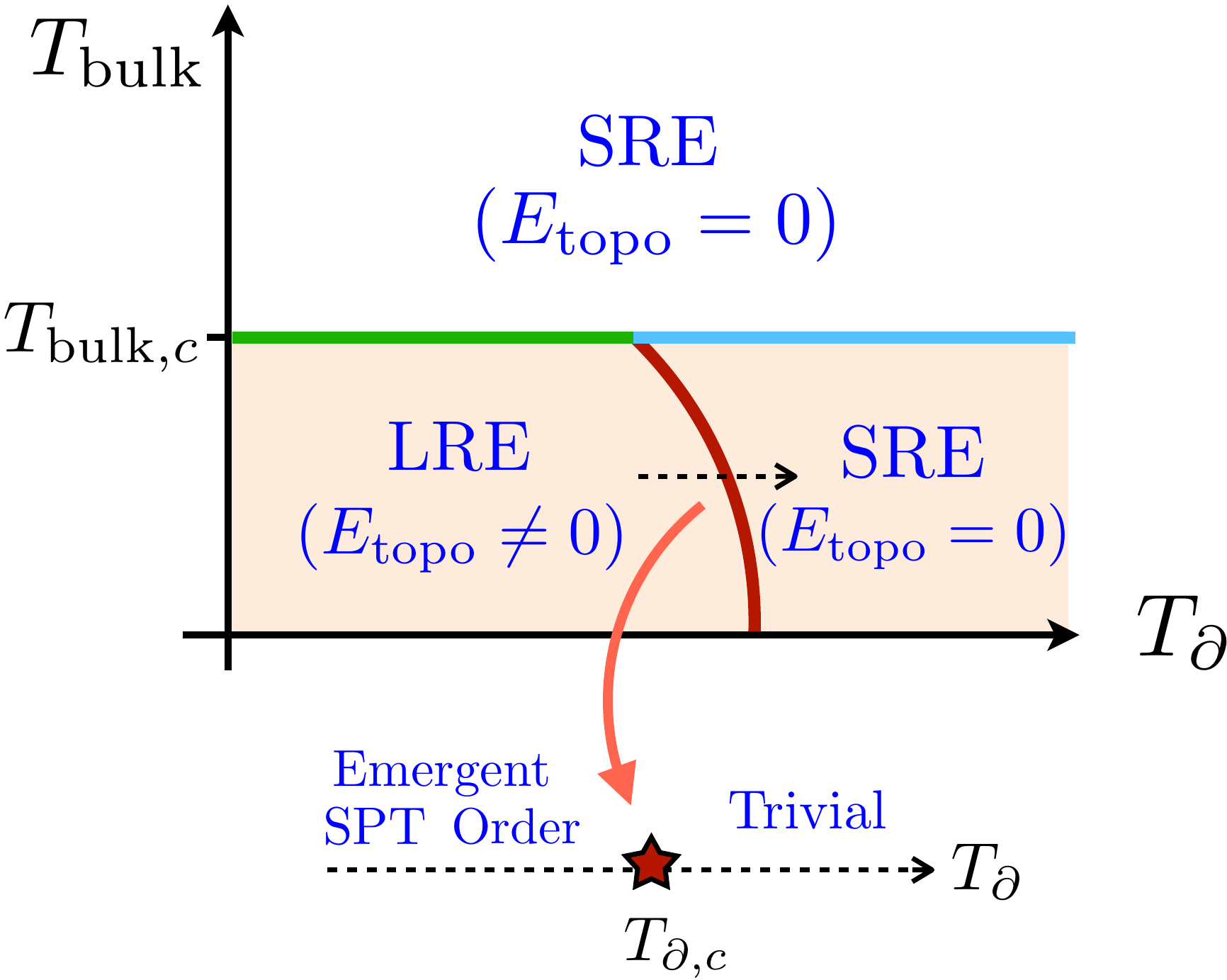}
	\end{subfigure}
	\caption{Schematic phase diagram for the topological negativity ($E_{\mathrm{topo}}$) of a thermal state with tunable temperatures in the bulk ($T_{\mathrm{bulk}}$) and bipartition boundary ($T_{\partial}$) in the 3d toric code when forbidding point-like excitations as well as 4d toric code when forbidding either type of excitations. When $T_{\textrm{bulk}}<T_{\text{bulk},c}$ the bulk is topologically ordered, and tuning $T_{\partial }$ drives a ``disentangling" phase transition where long-range entanglement across the bipartitioning surface is destroyed. This transition is related to the zero-temperature phase transition between an SPT order and a trivial order.  
	}  \label{fig:phase_diagram}
\end{figure}

\textit{\textbf{Entanglement negativity}}---
Given a density matrix $\hat{\rho}$ acting on a bipartite Hilbert space $\mathcal{H}= \mathcal{H}_{\mA}\otimes \mathcal{H}_{\mB}$, the entanglement negativity between $\mA$ and $\mB$ is defined by taking the partial transpose of the density matrix with respect to the Hilbert space of the $\mA$ subsystem. The negativity $E_N$ is defined with respect to this partially-transposed density matrix $\hat{\rho}^{T_{\mA}}$ as $E_N= \log\left(  \norm{ \hat{\rho}^{T_{\mA}}}_1  \right)$, where $\norm{M}_1$ denotes the 1-norm of the matrix $M$, i.e. the sum of all  absolute eigenvalues of $M$. 
%
%
 Given a local Hamiltonian $\hat{H}$ and a thermal density matrix $\hat{\rho}\sim e^{-\beta \hat{H}}$, the negativity $E_N$ between a subsystem $\mA$ of linear size $L_\mA$ and its complement $\mB$ in $d$ space dimensions can be written as $E_N= E_{\text{local}}-  E_{\text{topo}}$ \cite{lu2019structure,Lu_topo_nega_2020}. $E_{\text{local}}$ captures short-range entanglement along the bipartition boundary, and exhibits area-law scaling $E_{\text{local}} 
 \sim  \alpha_{d-1 } L_{\mA}^{d-1}$ to leading order in $L_{\mA}$ while  
 $E_{\text{topo}}$ is the \emph{topological entanglement negativity}, a universal constant contribution that is believed to characterize the long-range entanglement in topological order.

\textit{\textbf{Entanglement Negativity at Zero Temperature and an Emergent SPT Order}}--- We now demonstrate the  emergence of an SPT order on the entanglement bipartition in a topologically-ordered state at zero temperature, and the relevance of this order for the topological entanglement negativity. We focus on the $\mathbb{Z}_{2}$ topological order in various spatial dimensions, and find that the emergent SPT order is protected by a $\mathbb{Z}_{2}\cross \mathbb{Z}_{2}$ symmetry. 
 The SPT order hosts two distinct symmetry charges  corresponding to two species of gapped excitations in $\mathbb{Z}_{2}$ topological order, namely the charge ($e$) and flux ($m$), and the protecting symmetry is given by the action of emergent conservation laws in the topological order along the entanglement bipartition. 
The emergent SPT order completely characterizes the negativity spectrum (i.e. the eigenspectrum of the partially transposed density matrix $\hat{\rho}^{T_{\mA}}$) in that various eigenvalues correspond to various operator choices in the strange correlators\cite{You_strange_correlator_2014}  that diagnoses the SPT orders. 

Here we outline the approach for taking a partial transpose and discuss the emergence of SPT wave functions localized on the bipartition boundary. Consider a stabilizer Hamiltonian $\hat{H}= -J \sum_{j=1}^N \hat{\theta}_j$, where each stabilizer $ \hat{\theta}_j$ is a tensor product of Pauli operators over lattice sites. Distinct terms in the Hamiltonian mutually commute, and each stabilizer has eigenvalues $\theta_{j} = \pm 1$.  As a result, the density matrix for this system at zero temperature is $\hat{\rho} = \prod_{ j} \frac{1+\hat{\theta}_j}{2} = 2^{-N}\sum_{ \{ s_j  \}}  \left[ \prod_j \hat{\theta}_j^{s_j }  \right]$ with $s_j\in\{0,1\}$. 
Alternatively, the spectrum $\rho$ can be expressed as an expectation value of operators evaluted in the Hilbert space of the $\{s_j\}$ variables: $\rho  =  \bra{ +} \prod_j Z_j^{\frac{1-\theta_j}{2}}  \ket{ +}$, where $\ket{+} =2^{-N/2} \sum_{\{s_j\}}\ket{ \{s_j\} }  $, and the Pauli operator $Z_{j}$ acts within the $\{s_{j}\}$ Hilbert space as $Z_{j}|s_{j}\rangle = (1-2s_{j})|s_{j}\rangle$. Only the choice of stabilizer eigenvalues $\theta_j =1 ~\forall ~j$ gives a non-zero  eigenvalue of the density matrix, as expected.

We now consider the entanglement negativity within the ground state. By dividing the system into disjoint subsystems $\mA$ and $\mB$, taking a partial transpose on $\mA$ gives: $\left[   \prod_j   \hat{\theta}_j^{s_j } \right]^{T_{\mA}} = \left[   \prod_j   \left(\hat{\theta}_j^{T_\mA}\right)^{s_j } \right] \psi(s  )$, where $\psi(s) = \pm 1$ is a non-trivial sign determined by the number of pairs of stabilizers that anticommute when restricted in $\mA$. Specifically, introducing  $\hat{\theta}_i  |_\mA$ to denote the part of $\hat{\theta}_i$ that acts within $\mA$ and the matrix $C$ that encodes the commutation relation among $\hat{\theta}_i |_{\mA}$:  $C_{ij} =0, 1 $ for $[ \hat{\theta}_i|_\mA , \hat{\theta}_j |_\mA] =0$ and  $\{ \hat{\theta}_i|_\mA , \hat{\theta}_j |_\mA\} =0$, respectively, the sign introduced by partial transpose is 
\begin{equation}\label{eq:sign}
\psi (  \{ s_j   \} )  =  (-1)^{  \sum_{i<j } s_i C_{ij}s_j  }.
\end{equation}

We now focus on the stabilizer Hamiltonian for the toric code, which describes $\mathbb{Z}_{2}$ topological order in $d$ spatial dimensions.  Since only those stabilizers on the bipartition boundary can anticommute with each other when restricted in $\mA$, the negativity spectrum of the toric code at zero temperature depends only on the boundary part of the Hamiltonian, namely,  $\hat{H}_{\partial}= -\lambda_A \sum_{i \in R_a} \hat{A}_i -\lambda_B \sum_{j\in R_b} \hat{B}_j $, where $\{\hat{A}_i\}$, $\{\hat{B}_j\}$ are the Pauli-$X$ and $Z$-type stabilizers corresponding to the gapped $\mathbb{Z}_{2}$ charge and flux excitations of the toric code, while $R_a$, $R_b$ denote the locations of those stabilizers acting across the bipartitioning boundary. Following Eq.\ref{eq:sign}, one finds $\left[\prod_{i \in R_a}\hat{A}_{i}^{a_{i}} \prod_{j \in R_b}\hat{B}_{j}^{b_{j}}\right]^{T_{\mA}} = \left[\prod_{i \in R_a}\hat{A}_{i}^{a_{i}} \prod_{j\in R_b}\hat{B}_{j}^{b_{j}}\right]\psi(a,b)$, with  $a_{i}$, $b_{j}\in\{0,1\}$ and the sign $\psi(a,b)$ may be written as
\begin{equation}\label{eq:spt_wf}
\psi(a,b) = \prod_{i \in R_a}(-1)^{a_i \sum_{ j\in \partial i  }b_j    }=   \prod_{j \in R_b}(-1)^{b_j \sum_{ i\in  \partial j  } a_i   },
\end{equation}
where $\sum_{j \in \partial i }b_j$ denotes a sum of $b_j$ variables adjacent to $a_i$, and $\sum_{ i\in  \partial j  } a_i $  denotes a sum over the $a_i$ variables adjacent to $b_j$. This is because any two adjacent $\hat{A}_i$ and $\hat{B}_j$ must anticommute when restricted on a subregion. Using Eq. \ref{eq:spt_wf}, one can define  the state $ \ket{\psi } = \frac{1}{\sqrt{\mathcal{N}}}\sum_{ \{a_i,b_j \} } \psi(a,b) \ket{a,b}$ with normalization constant $\mathcal{N}$, and the negativity spectrum can be derived as \cite{appendix}
\begin{equation}
\rho^{T_{\mA }}  =\bra{  +  }   \prod_{i \in R_a}  Z_i^{\frac{1- A_i}{2}}   \prod_{j\in R_b}  Z_j^{\frac{1- B_j}{2}}    \ket{  \psi}.
\end{equation}

The wavefunction $\ket{\psi}$ exhibits a non-trivial SPT order with respect to $\mathbb{Z}_{2} \cross \mathbb{Z}_{2} $ symmetries, which arise from restricting the conservation laws obeyed by the $\mathbb{Z}_{2}$ charge and flux excitations to the entanglement bipartition.  Each conservation law (e.g. the global $\mathbb{Z}_{2}$ charge conservation $\prod_{i}A_{i\in R_{a}} = 1$ for the $d=2$ toric code) gives rise to a symmetry of the wavefunction (e.g. $a_{i}\rightarrow 1-a_{i}$ on all sites).  In addition, $\ket{\psi}$ is the ground-state of the Hamiltonian $H_p = - \sum_{i   \in R_a} X_i \prod_{j \in \partial i }Z_i  - \sum_{ j \in R_b} X_j \prod_{i \in \partial j }Z_j   $, where the first term is the product of a Pauli-X on site $i$ in $R_a$ and the Pauli-Zs on its neighboring  sites in $R_b$. The second term is defined similarly. The SPT wave function $\psi(a,b)$ has a $\mathbb{Z}_{2} \cross \mathbb{Z}_{2} $ symmetry may be understood in a ``decorated domain wall'' picture \cite{chen_decorated}; the phase $\prod_{i\in R_a}(-1)^{a_i \sum_{ j\in \partial i  }b_j    }$ implies decorating the domain walls of $b_j$ charges using $a_i$ charges, and $ \prod_{j\in R_b}(-1)^{b_j \sum_{ i\in  \partial j  } a_i   } $ implies decorating the domain walls of $a_i$ charges using $b_j$ charges. 

The robust braiding of the symmetry defects in this SPT order can be observed in strange correlators, and gives rise to a topological entanglement negativity for the original $\mathbb{Z}_{2}$ toric code. Below we demonstrate  these features using the 2d toric code as an example. The model is defined on a 2d lattice with every bond accommodating a spin-$1/2$ degree of freedom. The Hamiltonian reads $\hat{H}= -  \sum_s \hat{A}_s - \sum_p \hat{B}_p  $, where $\hat{A}_s$ is the product of four Pauli-X's on bonds emanating from a site $s$, and $\hat{B}_p$ is the product of four Pauli-Z's on bonds locating on the boundary of a plaquette $p$. Considering a subsystem $\mA$ with a closed boundary of size $L$, one can label the star and plaquette stabilizers on the boundary as $\hat{A}_1, \hat{B}_1, \hat{A}_2, \hat{B}_2, \cdots , \hat{A}_L, \hat{B}_L$, and the negativity spectrum is $\rho^{T_\mA}   = \bra{  +  } \prod_{j=1}^{L} \left[Z_{2j-1}^{\frac{1-A_j   }{2}}  Z_{2j}^{\frac{1-B_j   }{2}}    \right] \ket{  \psi}$ where $\ket{\psi}$ is the 1d cluster state with the parent Hamiltonian $H_p = -\sum_{\text{odd}~i} Z_{i-1} X_{i }Z_{i+1}  -\sum_{\text{even}~ i} Z_{i-1} X_{i }Z_{i+1}$ under the periodic boundary condition. $\ket{\psi}$ exhibits an SPT order protected by the $\mathbb{Z}_{2}\cross \mathbb{Z}_{2}$ symmetry generated by $\prod_{\text{odd}~ i}X_i$ and $\prod_{\text{even} ~ i }X_i$. As a consequence, only even number of $A_j$ excitations ($A_j=-1$) and $B_j$ excitations ($B_j = -1$) can give non-vanishing strange correlators. This number parity conservation reflects the number parity conservation of anyon charges in the topological order. 


Key features in the entanglement negativity are encoded in the SPT order. For the eigenvalue of $\rho^{T_{\mA}}$ with two excitations $A_m=A_n=-1$, the corresponding strange correlator gives $\bra{  +  }  Z_{2m-1 }Z_{2 n-1}  \ket{  \psi} = \bra{  +  }  Z_{2m-1 }Z_{2 n-1}  \left[\prod_{i=m}^{n-1} Z_{2i-1}X_{2i} Z_{2i+1} \right] \ket{  \psi}= \bra{  +  }\ket{  \psi}$. 
On the other hand, exciting both $A_j$ and $B_j$ in such a way that the excitations must be exchanged in order to return to the ground state 
 produces a sign $(-1)$  for the strange correlator. A simple example is given by considering $A_1= B_1=A_2=B_2=-1$, where the corresponding eigenvalue $\bra{  +  }  Z_{1}Z_2Z_3Z_4  \ket{  \psi}$ can be written as $ 			\bra{  +  }  Z_{1}Z_2Z_3Z_4  \ket{  \psi} = \bra{  +  }  Z_{1}Z_2Z_3Z_4 (Z_1 X_2Z_3) ( Z_2 X_3Z_4   )   \ket{  \psi} =  - \bra{+}\ket{\psi}$. One can easily generalize the above discussion to arbitrary pattens of excitations, and conclude that operators $O$ which are invariant under the $\mathbb{Z}_{2}\times\mathbb{Z}_{2}$ symmetry exhibit order in the strange correlator $\bra{+}O\ket{\psi} \ne 0$, and that the spectrum of $\rho^{T_{\mA}}$ only contains two eigenvalues $\pm \bra{+}\ket{\psi}$, where plus/minus sign corresponds to trivial/non-trivial braiding of two species of excitations. Knowledge of the entire negativity spectrum allows us to derive the zero temperature entanglement negativity \cite{appendix} $E_N=L\log 2 -  E_{\textrm{topo}} $ with $E_{\textrm{topo}} = \log 2$ being the topological entanglement negativity that reflects the underlying topological order.  $E_{\textrm{topo}}$ may be understood as a universal reduction in the negativity due to the fact that only operators $O$ which are invariant under the $\mathbb{Z}_{2}\times\mathbb{Z}_{2}$ symmetry exhibit order in the strange correlator $\bra{+}O\ket{\psi} \ne 0$.

\textit{ \textbf{Entanglement Negativity Transition at Finite Temperature}}--- For stabilizer models at finite temperature, the partial transpose still acts non-trivially on the boundary of $\mA$ region, resulting in $\hat{\rho}^{T_{\mA}}  \sim e^{ -\beta (\hat{H}_\mA  +\hat{H}_\mB  )} (e^{-\beta \hat{H}_{\partial}})^{T_\mA}$, where $\hat{H}_{\mA}/\hat{H}_{\mB}$ contains stabilizers supported only on $\mA/\mB$, and $\hat{H}_{\partial}$ denotes the interaction between $\mA$ and $\mB$. The negativity spectrum from the boundary part remains in the form of a strange correlator: $(e^{-\beta \hat{H}_{\partial}})^{T_\mA} \sim \bra{  +  }   \prod_{i \in R_a}  Z_i^{\frac{1- A_i}{2}}   \prod_{j\in R_b}  Z_j^{\frac{1- B_j}{2}}    \ket{  \psi(T)}$, but crucially, a non-zero temperature $T$ amounts to introducing a symmetry-breaking field that tends to polarize $a_i, b_j$ spins: $\ket{\psi (T)}\sim
 \sum_{\{a_i,b_j \} } \psi(a,b) \prod_{i \in R_a} \tanh(\beta \lambda_A)^{ a_i   } \prod_{j \in R_b}  \tanh(\beta \lambda_B)^{ b_j  }  \ket{a,b}$ where the phase $\psi(a,b)$ are given by Eq. \ref{eq:spt_wf}. The Hamiltonian whose ground-state is $\ket{ \psi (T)}$ is given by
 \begin{equation}
 	\begin{split}
 	H_p= & -  \sum_{i\in R_a } [ X_i \prod_{j\in \partial i }  Z_i   + \sinh(K_A) Z_i     ] \\
 	&-  \sum_{j\in R_b } [ X_j \prod_{i\in \partial j }  Z_j   + \sinh(K_B) Z_j     ]    
 	\end{split}
 \end{equation} 
with $K_A = -\log[  \tanh(\beta \lambda_A)]  $ and $K_B = -\log[  \tanh(\beta \lambda_B)]  $, and therefore, any non-zero $ T/\lambda_A,  T/ \lambda_B$ correspond to the non-zero on-site field $K_A, K_B$, as shown in the Supplemental Material \cite{appendix}.

For the thermal density matrix of the 2d toric code under partial transposition, the corresponding state $\ket{\psi(T)}$ on the entangling surface is a 1d cluster state purturbed by an onsite field, which destroys the SPT order. This can be seen by computing the negativity spectrum, which takes the form of correlation functions in 1d Ising model, with  different choices of $\{A_i\}$ and $\{ B_i \}$ corresponding to different spin insertion and coupling strength between neighboring spins: $ \left(  e^{-\beta H_{\partial}} \right)^{T_\mA} \sim \sum_{  \{ \tau_i =\pm 1  \}    } \prod_{i} \tau_i^{\frac{1-A_i }{2}} e^{ -K_A \sum_i \frac{1-\tau_i}{2}  +  \beta \lambda_B\sum_{i } B_i \tau_i \tau_{i+1}    }$.  Since the Ising order cannot survive at any finite temperature or any finite symmetry-breaking field, the negativity spectrum is short-range correlated, and the non-trivial braiding structure between $\{A_i\}$ and $\{B_i\}$ no longer exists. As a result, $E_{\mathrm{topo}}=0$ at any non-zero temperature in the thermodynamic limit\cite{Lu_topo_nega_2020}, corresponding to the vanishing topological order in the thermal Gibbs state.

\textbf{\textit{3d toric code}}  --- Now we discuss the negativity spectrum and the emergent SPT wave function localized on the entangling surface for the 3d toric code with the Hamiltonian $H= -\sum_s \hat{A}_s  - \sum_p \hat{B}{}_p$, where $\hat{A}_s$ is the product of six Pauli-Xs on bonds emanating from a site $s$, and $\hat{B}_p$ is the product of four Pauli-Z on bonds on the boundary of a plaquette $p$. Therefore, the stabilizers on the 2d bipartition boundary consists of $\hat{A}_i$ living on sites and $\hat{B}_{ij}$ living on links. Using the formalism introduced above, the negativity spectrum of $\hat{\rho}^{T_{\mA}}$ at zero temperature reads $	\rho^{T_\mA}   = \bra{  +  }\left[ \prod_{i}Z_i^{\frac{1- A_i}{2}}   \right] \left[\prod_{\expval{ij}}Z_{ij}^{\frac{1- B_{ij}}{2}}  \right] \ket{  \psi}$, where $\ket{\psi}$ is the ground state of the Hamiltonian $H_p= - \sum_i X_i \prod_{j \in\partial i }  Z_{ij} - \sum_{\expval{ij}    }  Z_i X_{ij}Z_j $. The first term $X_i \prod_{j \in\partial i }  Z_{ij} $ is a product of Pauli-X on the site $i$ and four Pauli-Zs acting on links whose boundary contains the site $i$. The second term $Z_i X_{ij}Z_j $ is a product of Pauli-X on the link $\expval{ij}$ and two Pauli-Zs on the boundary of the link. $\ket{\psi}$ exhibits an SPT order protected by a $\mathbb{Z}_{2}$ 0-form $\cross$ $\mathbb{Z}_{2}$ 1-form symmetry. Here the 0-form symmetry is implemented by $\prod_i X_i $ and the  1-form symmetry is implemented by $\prod_{\expval{ij} \in \mathcal{C}} X_{ij}$, where $\mathcal{C}$ is any 1d closed loop. Note that such an SPT order defined on a two-dimensional three-colorable graph has been discussed in Ref.\cite{yoshida_2016_symmetry}. 

%
%

Due to the $\mathbb{Z}_{2}$ 0-form and $\mathbb{Z}_{2}$ 1-form symmetry, eigenvalues of $\rho^{T_\mA}$ are non-vanishing only when the number of excitations in $A_i$ is even, and excitations in $B_{ij}$ exist  along a closed loop. Similar to the 2d toric code at zero temperature, the negativity spectrum only contains two distinct eigenvalues with an opposite sign $\pm \bra{+}\ket{\psi}$, where the minus sign corresponds to the non-trivial braiding between two species of excitations. Specifically, a loop formed by $B_{ij}$ excitations enclosing an odd number of $A_i$ excitation will give a minus one sign. In this case, the degeneacy of negativity spectrum originates from the long-range correlation in the following two kinds of strange correlators:  $\bra{  +  }  Z_{i}Z_j   \ket{  \psi} / \bra{  +  }\ket{  \psi} =1 $ and $\bra{  +  }   \prod_{\expval{ij}  \in \mathcal{C}  }  Z_{ij } \ket{  \psi} / \bra{  +  }\ket{  \psi} =1 $.

At a finite temperature, the boundary part of the negativity spectrum is $\left( e^{-\beta H_{\partial}} \right)^{T_{\mA}} \sim \bra{+} \prod_{i}Z_i^{\frac{1-A_i}{2}}  \prod_{\expval{ij}}   Z_{ij}^{\frac{1-B_{ij}}{2}} \ket{\psi(T)} $, where $\ket{\psi(T)}$ is the ground state of the perturbed SPT Hamiltonian $H_p= - \sum_i [ X_i \prod_{j \in\partial i }  Z_{ij}  + \sinh(K_A) Z_i  ]    - \sum_{\expval{ij}    }   [ Z_i X_{ij}Z_j  + \sinh(K_B) Z_{ij} ]$. The strange correlators can be written as correlation functions in the 2d Ising model: 

\begin{equation}
     \sum_{ \{\tau_i \} }  \prod_i \tau_i^{\frac{1-A_i}{2}} e^{ -K_A \sum_{i} \frac{1-\tau_i}{2}   +  \beta \lambda_B \sum_{\expval{ij}   } B_{ij} \tau_i \tau_j   }.
\end{equation}
As the 2d Ising model exhibits a spontaneous symmetry-breaking order up to a finite critical temperature $T_c$ only in the absence of symmetry-breaking fields (i.e. $K_A =0$), the existence of the long-range strange correlator and the long-range entanglement negativity requires taking  $\beta \lambda_A\to \infty$. This is consistent with the observation that forbidding point-like excitations in the Gibbs state of 3d toric code supports a finite-T topological order\cite{yoshida2011,hamma2012topological,castelnovo2008topological,Lu_topo_nega_2020}. Within the SPT order picture, setting $\beta \lambda_A\to \infty$ while tuning $\beta \lambda_B$ corresponds to enforcing the zero-form symmetry while adding a perturbation to break the 1-form symmetry. Crucially, the 1-form symmetry will be emergent at low energy as long as the Hamiltonian remains gapped and away from a quantum critical point as shown by Hastings and Wen based on the  quasi-adiabatic continuation\cite{hastings_2005_gauge}. Therefore, the boundary SPT order is  protected by this emergent symmetry up to a finite perturbation strength, corresponding to the persistence of long-range entanglement  up to a finite critical temperature.

When prohibiting the excitations away from the entangling surface and only allowing the loop-like excitations on the entangling surface, the transition in long-range entanglement negativity corresponds to the transition from an SPT order to a trivial state, which turns out to be the order-disorder transition in the 2d Ising model. Specifically, we derive the exact entanglement negativity $E_N$ in the 3d toric code with a $L\times L$ bipartition boundary at all temperatures in the Supplemental Material \cite{appendix}:
%
%

\begin{equation}\label{eq:negativity}
E_N(T) = L^2 \log 2 - \beta E_g   -\log Z(T), 
\end{equation}
where  $E_g =-2L^2 \lambda_B $  is the ground state energy and $Z(T)= \sum_{\{\tau_i  \}} e^{ \beta \sum_{\expval{ij}} \tau_i\tau_j   }$ is the partition function for the  Ising model on a square lattice of size $L\times L$. This result indicates that the negativity relates to  the free energy in the 2d  Ising model. As approaching the zero temperature, the partition function $Z$ take the form of $N_g e^{-\beta E_g}$ with $N_g=2$ being the number of ground states, giving the zero-temperature negativity $ E_N(T=0) =L^2  \log 2 -  \log N_g $. As a result, one finds topological entanglement negativity $E_{\text{topo}}$ simply reveals the number of symmetry broken sectors through the expression $E_{\text{topo}}=  \log N_g = \log 2$. Moreover, since the transition in the Ising model occurs at a finite temperature $T_c$, above which $\log Z(T)$ no longer has the subleading term $\log N_g$ in the thermodynamic limit (due to the restoration of the Ising symmetry), the topological entanglement negativity exhibits a discontinuity in the thermodynamic limit:  $E_{\text{topo}}= \log 2 ,  0 $ for $T< T_c$ and $T>T_c$, corresponding to the presence and  absence of long-range entanglement  across the bipartition surface. In particular, as $T\to T_c^+$, the Ising partition function is dominated by the largest and the next-largest eigenvalues of the row transfer matrix, i.e.  $Z\approx \lambda_0^L+ \lambda_1^L = \lambda_0^L (  1+  ( \lambda_1/\lambda_0 )^L ) \approx \lambda_0^L (1+ e^{ -L/\xi }) $  with $\xi $ being the correlation length for the 2-points function in the 2d Ising model\cite{baxter2016exactly}. This suggests the following scaling form of topological negativity
\begin{equation}\label{eq:3dtoric_scaling}
	E_{\text{topo}}= \log \left(   1+ e^{ -L/ \xi }  \right).
\end{equation}
where the critical properties in $\xi$ are within the 2d Ising universality class.


Above we have discussed a disentangling transition in long-range entanglement is destroyed by thermalizing the boundary while the bulk remains fixed at zero temperature. Now we consider  the general situation (Fig. \ref{fig:phase_diagram}) with a tunable bulk temperature  $T_{\textrm{bulk}}= 1/\beta_{\textrm{bulk}}$ and a tunable boundary temperature $T_{\partial}= 1/\beta_{\partial}$, namely, $\hat{\rho}\sim e^{  -\beta_{\textrm{bulk}} ( \hat{H}_{\mA} + \hat{H}_{\mB}   )} e^{ -\beta_{ \partial }  \hat{H}_{\partial}   }$. We still impose the condition $\lambda_A\to\infty$ so that the point-like excitations are prohibited in the thermal state. As $T_{\textrm{bulk}}< T_{ \textrm{bulk,c}}$, the bulk is topologically ordered, and the corresponding loop-like excitations are well-defined and obey an emergent, local constraint after an appropriate coarse-graining.  
This local constraint 
 results in an emergent one-form symmetry localized on the entanglement bipartition and gives rise to an emergent SPT order in the description of the partially-transposed density matrix; this implies that the topological negativity $E_{\text{topo}}=\log2$ at small $T_{\partial}$. More precisely, we show that the negativity relates to the annealed average of the 2d boundary theory over the 3d bulk fluctuation, and for $T_{\textrm{bulk}}< T_{\textrm{bulk,c}}$, the bulk fluctuations can be integrated out, leaving behind a coarse-grained 2d boundary theory whose universal properties remain the same as in $T_{\textrm{bulk}}=0$\cite{appendix}. Therefore, at a fixed  $T_{\textrm{bulk}}< T_{ \textrm{bulk,c}}$, increasing the boundary temperature $T_{\partial}$ leads to a disentangling transition from long-range entanglement ($E_{\text{topo}} = \log 2$) to short-range entanglement ($E_{\text{topo}} =  0$) where the universality belongs to the aforementioned 2d Ising universality.  Alternatively, the long-range entanglement can vanish by destroying the bulk topological order when increasing $T_{\text{bulk}}$ at a fixed $T_{\partial}$. This is because for $T_{\text{bulk}} > T_{\text{bulk,c}}$, the loop-like excitations are in the confined phase where the emergent gauge symmetry no longer exists. This in turn invalids the description  of SPT localized on the bipartition boundary, which indicates the absence of long-range entanglement.



\textbf{\textit{4d toric code}}  --- Finally, we discuss the 4d toric code where spins reside on each face (i.e. 2-cell) of a 4 dimensional hypercube. The Hamiltonian is $\hat{H}= -  \lambda_A \sum_l \hat{A}_l -  \lambda_B\sum_c \hat{B}_c$, where $\hat{A}_l$ is the product of 6 Pauli-X operators on the faces adjacent to the link $l$, and $\hat{B}_c$ is the product of 6 Pauli-Z operators on the faces around the boundary of the cube $c$. Since this model only possesses loop-like excitations, it exhibits a topological order up to a finite critical temperature. The boundary of a 4d hypercube is a 3d  lattice, where the boundary stabilizers are $\hat{A}_l$ living on links and $\hat{B}_p$ living on plaquettes. The negativity spectrum of 4d toric code at zero temperature reads $	\rho^{T_\mA}  = \bra{+}   \left[ \prod_{l} Z_l^{ \frac{1-A_l}{2}}   \right]  \left[ \prod_{p} Z_p^{ \frac{1-B_p}{2}}   \right]  \ket{\psi}$, where  $\ket{\psi}$ is the ground state of 3d cluster state Hamiltonian\cite{3d_cluster_state_2005}: $H_p=-  \sum_{ p } X_p \prod_{l: l \in \partial p  }Z_l  -  \sum_{ l } X_l \prod_{ p:  l \in \partial p   }      Z_p  $. The 3d cluster state exhibits an SPT order protected by the $\mathbb{Z}_{2}$ 1-form $\cross \mathbb{Z}_{2}$ 1-form symmetry so that the corresponding symmetry transformation acts on closed deformable two-dimensional surfaces. Specifically, any symmetry transformation can be obtained by taking the product of the following symmetry generators  $S_c=\prod_{p:p  \in\partial c  }X_p$ and $S_v= \prod_{l: v \in \partial l   }X_l$. It follows that the non-zero eigenvalules of $\rho^{T_\mA}$ require the operator $\left[ \prod_{l} Z_l^{ \frac{1-A_l}{2}}   \right]  \left[ \prod_{p} Z_p^{ \frac{1-B_p}{2}}   \right]  $ respect the one-form symmetry. This implies that these operators are  closed loops supported on the edges of the direct lattice and its dual lattice, and the braiding between a loop in the direct lattice and a loop in the dual lattice gives a $-1$ sign factor in the negativity spectrum as a consequence of the emergent SPT order. Note that the sign structure of braiding between loops in this SPT order has been discussed in Ref.\cite{Chong_2015_loop}.

At finite temperature, negativity spectrum from the boundary Gibbs state is given by the strange correlator  $\left( e^{-\beta H_{\partial}} \right)^{T_{\mA}} \sim \bra{+} \left[ \prod_{l} Z_l^{ \frac{1-A_l}{2}}   \right]  \left[ \prod_{p} Z_p^{ \frac{1-B_p}{2}}   \right] \ket{\psi(T)}$, where $\ket{\psi (T)}$ is the ground state of $H_p=-  \sum_{ p } \left[  X_p \prod_{l: l \in \partial p  }Z_l  + \sinh(K_B) Z_p \right] -  \sum_{ l } \left[X_l \prod_{ p:  l \in \partial p   }      Z_p  +\sinh(K_A) Z_l\right] $, namely, the 3d cluster state Hamiltonian under an on-site perturbation. One finds that the strange correlators are the  Wilson loop operators in 3d Ising gauge theory coupled to dynamical matter fields

\begin{equation}\label{eq:4d_strange}
	\sum_{ \{\tau_i \} }  \prod_{l} \tau_l^{\frac{1-A_l}{2}} e^{ -K_A \sum_{l} \frac{1-\tau_l}{2}   +  \beta \lambda_B \sum_{ p  } B_{p}  \prod_{ l \in \partial p  }\tau_p     }
\end{equation}
with $K_A = -\log [ \tanh(\beta \lambda_A)]$. Since the deconfined phase of the gauge theory persists up to a finite $T/\lambda_A$, $T/\lambda_B$, \cite{fradkin_shenker_diagram,gauge_mc_calculation_Jayaprakash,gauge_phase_diagram_Stamp,gauge_phase_diagram_vidal,nahum_dual_2020}, the SPT order in $\ket{\psi(T)}$ persists up to a finite symmetry-breaking field that corresponds to $T_c$, below which the long-range entanglement negativity exists. Note that despite the perturbation, the SPT order is protected by the emergent 1-form symmetry through out the entire deconfined phase\cite{hastings_2005_gauge,Wen_higer_symmetries_2019,nahum_dual_2020,McGreevy_one_form_2021}.


Using the emergent SPT order, we now discuss the nature of transition in long-range entanglement. As considering zero temperature in the bulk, tuning the boundary temperature drives a disentangling transition that corresponds to the presence/absence of long-range entanglement across the bipartition surface. The universal properties of this  transition is governed by the transition from an SPT order to a trivial state. In particular, for the case where one type of excitations on the entangling surface is prohibited, e.g. say $\beta \lambda_A\to \infty$ so the matter fields are absent in Eq.\ref{eq:4d_strange}, we determine the entanglement negativity in the Supplemental Material to be \cite{appendix}
\begin{equation}
E_N= 3L^3 - \beta E_g  - \log Z(T).
\end{equation}
Such an expression resembles Eq.\ref{eq:negativity} for the 3d toric code with point-like charges forbidden, but here $E_g=-3 L^3 \lambda_B$ and $Z(T)= \sum_{ \{ \tau_l \}  }  e^{ \beta \sum_p \prod_{ l \in \partial  p}  \tau_{ l }  }$ denote the ground state energy and the partition function in the 3d classical pure $\mathbb{Z}_2$ gauge theory. This expression implies that the transition in negativity is mapped to a confinement-deconfinement transition of the $\mathbb{Z}_{2}$ gauge theory. Moreover, across the critical temperature $T_c$ in the thermodynamic limit, $\log Z$ exhibits a discontinuity in its universal subleading term, i.e. $\log Z(T_c^- )  -\log Z(T_c^+ ) =2\log 2$, indicating  $E_{\text{topo}}= 2\log 2 ,  0 $ respectively for $T<T_c$ and $T>T_c$. This can be seen by mapping the finite-temperature 3d classical $\mathbb{Z}_{2}$ gauge theory to the ground subspace of $2+1$D quantum $\mathbb{Z}_{2}$ gauge theory $H= - \sum_p \prod_{ \expval{ij}\in \partial p  }  Z_{ij} - g \sum_{\expval{ij}} X_{ij}   $ with the Gauss law imposed on every vertex $\prod_{\expval{ij }\in +   }X_{ij}=1$. Tuning $g$ drives  a transition from a deconfined phase with four degenerate ground states to a confined phase with a single ground state,  thus resulting in a discontinuity of $\log 4$ in $\log Z$ at the critical point of the 3d classical $\mathbb{Z}_{2}$ gauge theory, which belongs to the 3d Ising universality.


Finally, for the more general case when the bulk is thermal (but one type of excitations is still prohibited), the schematic phase diagram of bulk topological order and the long-range entanglement still follows Fig.\ref{fig:phase_diagram}. In particular, when the bulk is topologically ordered ($T_{\textrm{bulk}} < T_{\textrm{bulk,c}}$), tuning the boundary temperature drives the disentangling transition, where the critical properties of long-range entanglement are still governed by the SPT-trivial order transition as we argue in the Supplemental Material\cite{appendix}.  On the other hand, for the case where both types of excitations are allowed on the boundary, since  the  negativity spectrum is described by 3d Ising gauge theory coupled to matter, we expect the transition in the negativity remains governed by a deconfinement transition. This is also suggested based on the replica calculation\cite{appendix}, but we are unable to derive a closed-form expression of negativity.


\textit{\textbf{Summary and discussion}}--- In this work, we point out an intriguing connection between topological order and SPT order via partial transpose. The gapped excitations in  $\mathbb{Z}_{2}$ topological order manifest as symmetry charges in the SPT order localized on the entangling surface, and these symmetry charges exhibit long-range correlation and robust braiding structure which reflect the underlying topological order. In particular, stability of topological order at finite temperature corresponds to stability of SPT order under symmetry breaking fields. The robustness of the SPT is possible if the broken symmetry is a one-form symmetry, in agreement with the fact that a finite-T topological order is allowed when supporting loop-like excitations. This provides a novel understanding in the existence of finite-T topological order in the 4d toric code, as well as the 3d toric code with point-like charges forbidden. In addition, assuming the excitations only occur on the bipartition boundary, we completely determine the nature of the transition for topological order via a mapping to certain statistical models.

We mainly focus on the $\mathbb{Z}_2$ topological order, so it is natural to ask whether the emergent SPT order picture applies to more general types of topological order. In this regard, we briefly discuss two other  generalizations, the details of which will be presented in the forthcoming work \cite{New_Paper}. First, our result can be generalized from $\mathbb{Z}_{2}$ to  $\mathbb{Z}_n$ gauge group, in which case a partial transpose acting on a $\mathbb{Z}_n$ topological order leads to a $\mathbb{Z}_n \cross \mathbb{Z}_n$ SPT order localized on the entangling surface. Such a calculation is more involved since a partial transpose acting on a stabilizer string with $\mathbb{Z}_n$ structure not only induces a non-trivial sign, but also acts non-trivially on stabilizers such that the resulting local operators no longer commute. This non-commuting structure means that the partially transposed Gibbs state is not  diagonal in the eigenbases of stabilizers. Nevertheless, SPT order is encoded in the matrix elements, and one can still analytically solve for the negativity spectrum. Second, the physics of emergent SPT orders induced by partial transpose is applicable to  fracton topological order. Specifically, for X-cube model\cite{xcube} and Haah's code\cite{Haah}, i.e. the representative of type-I and type-II fracton orders, a partial transpose results in an SPT order that is protected by subsystem symmetries\cite{you_2018_subsystem} and fractal symmetries respectively\cite{fractal_spt_2019_sondhi}.

\emph{Acknowledgments--} T.-C. Lu thanks Tarun Grover, Chong Wang, Liujun Zou for helpful discussions, and acknowledges support from Perimeter Institute for Theoretical Physics. Research at Perimeter Institute is supported in part by the Government of Canada through the Department of Innovation, Science and Economic Development and by the Province of Ontario through the Ministry of Colleges and Universities. SV acknowledges that part of this work was performed at the Aspen Center for Physics, which is supported by National Science Foundation grant PHY-1607611.

\bibliography{v1bib}

\begin{thebibliography}{69}%
\makeatletter
\providecommand \@ifxundefined [1]{%
 \@ifx{#1\undefined}
}%
\providecommand \@ifnum [1]{%
 \ifnum #1\expandafter \@firstoftwo
 \else \expandafter \@secondoftwo
 \fi
}%
\providecommand \@ifx [1]{%
 \ifx #1\expandafter \@firstoftwo
 \else \expandafter \@secondoftwo
 \fi
}%
\providecommand \natexlab [1]{#1}%
\providecommand \enquote  [1]{``#1''}%
\providecommand \bibnamefont  [1]{#1}%
\providecommand \bibfnamefont [1]{#1}%
\providecommand \citenamefont [1]{#1}%
\providecommand \href@noop [0]{\@secondoftwo}%
\providecommand \href [0]{\begingroup \@sanitize@url \@href}%
\providecommand \@href[1]{\@@startlink{#1}\@@href}%
\providecommand \@@href[1]{\endgroup#1\@@endlink}%
\providecommand \@sanitize@url [0]{\catcode `\\12\catcode `\$12\catcode
  `\&12\catcode `\#12\catcode `\^12\catcode `\_12\catcode `\%12\relax}%
\providecommand \@@startlink[1]{}%
\providecommand \@@endlink[0]{}%
\providecommand \url  [0]{\begingroup\@sanitize@url \@url }%
\providecommand \@url [1]{\endgroup\@href {#1}{\urlprefix }}%
\providecommand \urlprefix  [0]{URL }%
\providecommand \Eprint [0]{\href }%
\providecommand \doibase [0]{http://dx.doi.org/}%
\providecommand \selectlanguage [0]{\@gobble}%
\providecommand \bibinfo  [0]{\@secondoftwo}%
\providecommand \bibfield  [0]{\@secondoftwo}%
\providecommand \translation [1]{[#1]}%
\providecommand \BibitemOpen [0]{}%
\providecommand \bibitemStop [0]{}%
\providecommand \bibitemNoStop [0]{.\EOS\space}%
\providecommand \EOS [0]{\spacefactor3000\relax}%
\providecommand \BibitemShut  [1]{\csname bibitem#1\endcsname}%
\let\auto@bib@innerbib\@empty
\bibitem [{\citenamefont {Wen}(1989)}]{wen1989vacuum}%
  \BibitemOpen
  \bibfield  {author} {\bibinfo {author} {\bibfnamefont {X.~G.}\ \bibnamefont
  {Wen}},\ }\href {\doibase 10.1103/PhysRevB.40.7387} {\bibfield  {journal}
  {\bibinfo  {journal} {Phys. Rev. B}\ }\textbf {\bibinfo {volume} {40}},\
  \bibinfo {pages} {7387} (\bibinfo {year} {1989})}\BibitemShut {NoStop}%
\bibitem [{\citenamefont {Wen}\ and\ \citenamefont
  {Niu}(1990)}]{wen1990ground}%
  \BibitemOpen
  \bibfield  {author} {\bibinfo {author} {\bibfnamefont {X.~G.}\ \bibnamefont
  {Wen}}\ and\ \bibinfo {author} {\bibfnamefont {Q.}~\bibnamefont {Niu}},\
  }\href {\doibase 10.1103/PhysRevB.41.9377} {\bibfield  {journal} {\bibinfo
  {journal} {Phys. Rev. B}\ }\textbf {\bibinfo {volume} {41}},\ \bibinfo
  {pages} {9377} (\bibinfo {year} {1990})}\BibitemShut {NoStop}%
\bibitem [{\citenamefont {Wen}(1990)}]{wen1990topological}%
  \BibitemOpen
  \bibfield  {author} {\bibinfo {author} {\bibfnamefont {X.~G.}\ \bibnamefont
  {Wen}},\ }\href {\doibase 10.1142/S0217979290000139} {\bibfield  {journal}
  {\bibinfo  {journal} {International Journal of Modern Physics B}\ }\textbf
  {\bibinfo {volume} {04}},\ \bibinfo {pages} {239} (\bibinfo {year}
  {1990})}\BibitemShut {NoStop}%
\bibitem [{\citenamefont {Dennis}\ \emph {et~al.}(2002)\citenamefont {Dennis},
  \citenamefont {Kitaev}, \citenamefont {Landahl},\ and\ \citenamefont
  {Preskill}}]{dennis2002}%
  \BibitemOpen
  \bibfield  {author} {\bibinfo {author} {\bibfnamefont {E.}~\bibnamefont
  {Dennis}}, \bibinfo {author} {\bibfnamefont {A.}~\bibnamefont {Kitaev}},
  \bibinfo {author} {\bibfnamefont {A.}~\bibnamefont {Landahl}}, \ and\
  \bibinfo {author} {\bibfnamefont {J.}~\bibnamefont {Preskill}},\ }\href
  {\doibase 10.1063/1.1499754} {\bibfield  {journal} {\bibinfo  {journal}
  {Journal of Mathematical Physics}\ }\textbf {\bibinfo {volume} {43}},\
  \bibinfo {pages} {4452} (\bibinfo {year} {2002})}\BibitemShut {NoStop}%
\bibitem [{\citenamefont {Nussinov}\ and\ \citenamefont
  {Ortiz}(2008)}]{nussinov2008autocorrelations}%
  \BibitemOpen
  \bibfield  {author} {\bibinfo {author} {\bibfnamefont {Z.}~\bibnamefont
  {Nussinov}}\ and\ \bibinfo {author} {\bibfnamefont {G.}~\bibnamefont
  {Ortiz}},\ }\href {\doibase 10.1103/PhysRevB.77.064302} {\bibfield  {journal}
  {\bibinfo  {journal} {Phys. Rev. B}\ }\textbf {\bibinfo {volume} {77}},\
  \bibinfo {pages} {064302} (\bibinfo {year} {2008})}\BibitemShut {NoStop}%
\bibitem [{\citenamefont {Nussinov}\ and\ \citenamefont
  {Ortiz}(2009)}]{nussinov2009sufficient}%
  \BibitemOpen
  \bibfield  {author} {\bibinfo {author} {\bibfnamefont {Z.}~\bibnamefont
  {Nussinov}}\ and\ \bibinfo {author} {\bibfnamefont {G.}~\bibnamefont
  {Ortiz}},\ }\href {\doibase 10.1073/pnas.0803726105} {\bibfield  {journal}
  {\bibinfo  {journal} {Proceedings of the National Academy of Sciences}\
  }\textbf {\bibinfo {volume} {106}},\ \bibinfo {pages} {16944} (\bibinfo
  {year} {2009})}\BibitemShut {NoStop}%
\bibitem [{\citenamefont {Bravyi}\ and\ \citenamefont
  {Terhal}(2009)}]{bravyi2009no_go}%
  \BibitemOpen
  \bibfield  {author} {\bibinfo {author} {\bibfnamefont {S.}~\bibnamefont
  {Bravyi}}\ and\ \bibinfo {author} {\bibfnamefont {B.}~\bibnamefont
  {Terhal}},\ }\href {\doibase 10.1088/1367-2630/11/4/043029} {\bibfield
  {journal} {\bibinfo  {journal} {New Journal of Physics}\ }\textbf {\bibinfo
  {volume} {11}},\ \bibinfo {pages} {043029} (\bibinfo {year}
  {2009})}\BibitemShut {NoStop}%
\bibitem [{\citenamefont {Hastings}(2011)}]{hastings2011}%
  \BibitemOpen
  \bibfield  {author} {\bibinfo {author} {\bibfnamefont {M.~B.}\ \bibnamefont
  {Hastings}},\ }\href {\doibase 10.1103/PhysRevLett.107.210501} {\bibfield
  {journal} {\bibinfo  {journal} {Phys. Rev. Lett.}\ }\textbf {\bibinfo
  {volume} {107}},\ \bibinfo {pages} {210501} (\bibinfo {year}
  {2011})}\BibitemShut {NoStop}%
\bibitem [{\citenamefont {Yoshida}(2011)}]{yoshida2011}%
  \BibitemOpen
  \bibfield  {author} {\bibinfo {author} {\bibfnamefont {B.}~\bibnamefont
  {Yoshida}},\ }\href {\doibase https://doi.org/10.1016/j.aop.2011.06.001}
  {\bibfield  {journal} {\bibinfo  {journal} {Annals of Physics}\ }\textbf
  {\bibinfo {volume} {326}},\ \bibinfo {pages} {2566 } (\bibinfo {year}
  {2011})}\BibitemShut {NoStop}%
\bibitem [{\citenamefont {Landon-Cardinal}\ and\ \citenamefont
  {Poulin}(2013)}]{Poulin2013}%
  \BibitemOpen
  \bibfield  {author} {\bibinfo {author} {\bibfnamefont {O.}~\bibnamefont
  {Landon-Cardinal}}\ and\ \bibinfo {author} {\bibfnamefont {D.}~\bibnamefont
  {Poulin}},\ }\href {\doibase 10.1103/PhysRevLett.110.090502} {\bibfield
  {journal} {\bibinfo  {journal} {Phys. Rev. Lett.}\ }\textbf {\bibinfo
  {volume} {110}},\ \bibinfo {pages} {090502} (\bibinfo {year}
  {2013})}\BibitemShut {NoStop}%
\bibitem [{\citenamefont {Brown}\ \emph {et~al.}(2016)\citenamefont {Brown},
  \citenamefont {Loss}, \citenamefont {Pachos}, \citenamefont {Self},\ and\
  \citenamefont {Wootton}}]{brown2016review}%
  \BibitemOpen
  \bibfield  {author} {\bibinfo {author} {\bibfnamefont {B.~J.}\ \bibnamefont
  {Brown}}, \bibinfo {author} {\bibfnamefont {D.}~\bibnamefont {Loss}},
  \bibinfo {author} {\bibfnamefont {J.~K.}\ \bibnamefont {Pachos}}, \bibinfo
  {author} {\bibfnamefont {C.~N.}\ \bibnamefont {Self}}, \ and\ \bibinfo
  {author} {\bibfnamefont {J.~R.}\ \bibnamefont {Wootton}},\ }\href {\doibase
  10.1103/RevModPhys.88.045005} {\bibfield  {journal} {\bibinfo  {journal}
  {Rev. Mod. Phys.}\ }\textbf {\bibinfo {volume} {88}},\ \bibinfo {pages}
  {045005} (\bibinfo {year} {2016})}\BibitemShut {NoStop}%
\bibitem [{\citenamefont {Lu}\ \emph {et~al.}(2020)\citenamefont {Lu},
  \citenamefont {Hsieh},\ and\ \citenamefont {Grover}}]{Lu_topo_nega_2020}%
  \BibitemOpen
  \bibfield  {author} {\bibinfo {author} {\bibfnamefont {T.-C.}\ \bibnamefont
  {Lu}}, \bibinfo {author} {\bibfnamefont {T.~H.}\ \bibnamefont {Hsieh}}, \
  and\ \bibinfo {author} {\bibfnamefont {T.}~\bibnamefont {Grover}},\ }\href
  {\doibase 10.1103/PhysRevLett.125.116801} {\bibfield  {journal} {\bibinfo
  {journal} {Phys. Rev. Lett.}\ }\textbf {\bibinfo {volume} {125}},\ \bibinfo
  {pages} {116801} (\bibinfo {year} {2020})}\BibitemShut {NoStop}%
\bibitem [{\citenamefont {Peres}(1996)}]{peres1996}%
  \BibitemOpen
  \bibfield  {author} {\bibinfo {author} {\bibfnamefont {A.}~\bibnamefont
  {Peres}},\ }\href {\doibase 10.1103/PhysRevLett.77.1413} {\bibfield
  {journal} {\bibinfo  {journal} {Phys. Rev. Lett.}\ }\textbf {\bibinfo
  {volume} {77}},\ \bibinfo {pages} {1413} (\bibinfo {year}
  {1996})}\BibitemShut {NoStop}%
\bibitem [{\citenamefont {Horodecki}\ \emph {et~al.}(1996)\citenamefont
  {Horodecki}, \citenamefont {Horodecki},\ and\ \citenamefont
  {Horodecki}}]{horodecki1996}%
  \BibitemOpen
  \bibfield  {author} {\bibinfo {author} {\bibfnamefont {M.}~\bibnamefont
  {Horodecki}}, \bibinfo {author} {\bibfnamefont {P.}~\bibnamefont
  {Horodecki}}, \ and\ \bibinfo {author} {\bibfnamefont {R.}~\bibnamefont
  {Horodecki}},\ }\href {\doibase
  https://doi.org/10.1016/S0375-9601(96)00706-2} {\bibfield  {journal}
  {\bibinfo  {journal} {Physics Letters A}\ }\textbf {\bibinfo {volume}
  {223}},\ \bibinfo {pages} {1 } (\bibinfo {year} {1996})}\BibitemShut
  {NoStop}%
\bibitem [{\citenamefont {Eisert}\ and\ \citenamefont
  {Plenio}(1999)}]{eisert99}%
  \BibitemOpen
  \bibfield  {author} {\bibinfo {author} {\bibfnamefont {J.}~\bibnamefont
  {Eisert}}\ and\ \bibinfo {author} {\bibfnamefont {M.~B.}\ \bibnamefont
  {Plenio}},\ }\href {\doibase 10.1080/09500349908231260} {\bibfield  {journal}
  {\bibinfo  {journal} {Journal of Modern Optics}\ }\textbf {\bibinfo {volume}
  {46}},\ \bibinfo {pages} {145} (\bibinfo {year} {1999})}\BibitemShut
  {NoStop}%
\bibitem [{\citenamefont {Vidal}\ and\ \citenamefont
  {Werner}(2002)}]{vidal2002}%
  \BibitemOpen
  \bibfield  {author} {\bibinfo {author} {\bibfnamefont {G.}~\bibnamefont
  {Vidal}}\ and\ \bibinfo {author} {\bibfnamefont {R.~F.}\ \bibnamefont
  {Werner}},\ }\href {\doibase 10.1103/PhysRevA.65.032314} {\bibfield
  {journal} {\bibinfo  {journal} {Phys. Rev. A}\ }\textbf {\bibinfo {volume}
  {65}},\ \bibinfo {pages} {032314} (\bibinfo {year} {2002})}\BibitemShut
  {NoStop}%
\bibitem [{\citenamefont {Audenaert}\ \emph {et~al.}(2002)\citenamefont
  {Audenaert}, \citenamefont {Eisert}, \citenamefont {Plenio},\ and\
  \citenamefont {Werner}}]{audenaert2002entanglement}%
  \BibitemOpen
  \bibfield  {author} {\bibinfo {author} {\bibfnamefont {K.}~\bibnamefont
  {Audenaert}}, \bibinfo {author} {\bibfnamefont {J.}~\bibnamefont {Eisert}},
  \bibinfo {author} {\bibfnamefont {M.}~\bibnamefont {Plenio}}, \ and\ \bibinfo
  {author} {\bibfnamefont {R.}~\bibnamefont {Werner}},\ }\href@noop {}
  {\bibfield  {journal} {\bibinfo  {journal} {Physical Review A}\ }\textbf
  {\bibinfo {volume} {66}},\ \bibinfo {pages} {042327} (\bibinfo {year}
  {2002})}\BibitemShut {NoStop}%
\bibitem [{\citenamefont {Eisler}\ and\ \citenamefont
  {Zimbor{\'{a}}s}(2015)}]{Eisler_2015}%
  \BibitemOpen
  \bibfield  {author} {\bibinfo {author} {\bibfnamefont {V.}~\bibnamefont
  {Eisler}}\ and\ \bibinfo {author} {\bibfnamefont {Z.}~\bibnamefont
  {Zimbor{\'{a}}s}},\ }\href {\doibase 10.1088/1367-2630/17/5/053048}
  {\bibfield  {journal} {\bibinfo  {journal} {New Journal of Physics}\ }\textbf
  {\bibinfo {volume} {17}},\ \bibinfo {pages} {053048} (\bibinfo {year}
  {2015})}\BibitemShut {NoStop}%
\bibitem [{\citenamefont {Nobili}\ \emph {et~al.}(2016)\citenamefont {Nobili},
  \citenamefont {Coser},\ and\ \citenamefont {Tonni}}]{Tonni_negativity_2015}%
  \BibitemOpen
  \bibfield  {author} {\bibinfo {author} {\bibfnamefont {C.~D.}\ \bibnamefont
  {Nobili}}, \bibinfo {author} {\bibfnamefont {A.}~\bibnamefont {Coser}}, \
  and\ \bibinfo {author} {\bibfnamefont {E.}~\bibnamefont {Tonni}},\ }\href
  {\doibase 10.1088/1742-5468/2016/08/083102} {\bibfield  {journal} {\bibinfo
  {journal} {Journal of Statistical Mechanics: Theory and Experiment}\ }\textbf
  {\bibinfo {volume} {2016}},\ \bibinfo {pages} {083102} (\bibinfo {year}
  {2016})}\BibitemShut {NoStop}%
\bibitem [{\citenamefont {Eisler}\ and\ \citenamefont
  {Zimbor\'as}(2016)}]{Eisler_2016_free_lattice}%
  \BibitemOpen
  \bibfield  {author} {\bibinfo {author} {\bibfnamefont {V.}~\bibnamefont
  {Eisler}}\ and\ \bibinfo {author} {\bibfnamefont {Z.}~\bibnamefont
  {Zimbor\'as}},\ }\href {\doibase 10.1103/PhysRevB.93.115148} {\bibfield
  {journal} {\bibinfo  {journal} {Phys. Rev. B}\ }\textbf {\bibinfo {volume}
  {93}},\ \bibinfo {pages} {115148} (\bibinfo {year} {2016})}\BibitemShut
  {NoStop}%
\bibitem [{\citenamefont {Bianchini}\ and\ \citenamefont
  {Castro-Alvaredo}(2016)}]{Bianchini_2016_free_boson}%
  \BibitemOpen
  \bibfield  {author} {\bibinfo {author} {\bibfnamefont {D.}~\bibnamefont
  {Bianchini}}\ and\ \bibinfo {author} {\bibfnamefont {O.~A.}\ \bibnamefont
  {Castro-Alvaredo}},\ }\href {\doibase
  https://doi.org/10.1016/j.nuclphysb.2016.10.016} {\bibfield  {journal}
  {\bibinfo  {journal} {Nuclear Physics B}\ }\textbf {\bibinfo {volume}
  {913}},\ \bibinfo {pages} {879 } (\bibinfo {year} {2016})}\BibitemShut
  {NoStop}%
\bibitem [{\citenamefont {Shapourian}\ \emph {et~al.}(2017)\citenamefont
  {Shapourian}, \citenamefont {Shiozaki},\ and\ \citenamefont
  {Ryu}}]{Shapourian2017}%
  \BibitemOpen
  \bibfield  {author} {\bibinfo {author} {\bibfnamefont {H.}~\bibnamefont
  {Shapourian}}, \bibinfo {author} {\bibfnamefont {K.}~\bibnamefont
  {Shiozaki}}, \ and\ \bibinfo {author} {\bibfnamefont {S.}~\bibnamefont
  {Ryu}},\ }\href {\doibase 10.1103/PhysRevB.95.165101} {\bibfield  {journal}
  {\bibinfo  {journal} {Phys. Rev. B}\ }\textbf {\bibinfo {volume} {95}},\
  \bibinfo {pages} {165101} (\bibinfo {year} {2017})}\BibitemShut {NoStop}%
\bibitem [{\citenamefont {Shapourian}\ and\ \citenamefont
  {Ryu}(2019)}]{Shapourian2018}%
  \BibitemOpen
  \bibfield  {author} {\bibinfo {author} {\bibfnamefont {H.}~\bibnamefont
  {Shapourian}}\ and\ \bibinfo {author} {\bibfnamefont {S.}~\bibnamefont
  {Ryu}},\ }\href {\doibase 10.1088/1742-5468/ab11e0} {\bibfield  {journal}
  {\bibinfo  {journal} {Journal of Statistical Mechanics: Theory and
  Experiment}\ }\textbf {\bibinfo {volume} {2019}},\ \bibinfo {pages} {043106}
  (\bibinfo {year} {2019})}\BibitemShut {NoStop}%
\bibitem [{\citenamefont {Calabrese}\ \emph {et~al.}(2012)\citenamefont
  {Calabrese}, \citenamefont {Cardy},\ and\ \citenamefont
  {Tonni}}]{calabrese2012}%
  \BibitemOpen
  \bibfield  {author} {\bibinfo {author} {\bibfnamefont {P.}~\bibnamefont
  {Calabrese}}, \bibinfo {author} {\bibfnamefont {J.}~\bibnamefont {Cardy}}, \
  and\ \bibinfo {author} {\bibfnamefont {E.}~\bibnamefont {Tonni}},\ }\href
  {\doibase 10.1103/PhysRevLett.109.130502} {\bibfield  {journal} {\bibinfo
  {journal} {Phys. Rev. Lett.}\ }\textbf {\bibinfo {volume} {109}},\ \bibinfo
  {pages} {130502} (\bibinfo {year} {2012})}\BibitemShut {NoStop}%
\bibitem [{\citenamefont {Coser}\ \emph {et~al.}(2014)\citenamefont {Coser},
  \citenamefont {Tonni},\ and\ \citenamefont {Calabrese}}]{tonni_quench_cft}%
  \BibitemOpen
  \bibfield  {author} {\bibinfo {author} {\bibfnamefont {A.}~\bibnamefont
  {Coser}}, \bibinfo {author} {\bibfnamefont {E.}~\bibnamefont {Tonni}}, \ and\
  \bibinfo {author} {\bibfnamefont {P.}~\bibnamefont {Calabrese}},\ }\href
  {\doibase 10.1088/1742-5468/2014/12/p12017} {\bibfield  {journal} {\bibinfo
  {journal} {Journal of Statistical Mechanics: Theory and Experiment}\ }\textbf
  {\bibinfo {volume} {2014}},\ \bibinfo {pages} {P12017} (\bibinfo {year}
  {2014})}\BibitemShut {NoStop}%
\bibitem [{\citenamefont {Kulaxizi}\ \emph {et~al.}(2014)\citenamefont
  {Kulaxizi}, \citenamefont {Parnachev},\ and\ \citenamefont
  {Policastro}}]{negativity_large_c_2014_kulaxizi}%
  \BibitemOpen
  \bibfield  {author} {\bibinfo {author} {\bibfnamefont {M.}~\bibnamefont
  {Kulaxizi}}, \bibinfo {author} {\bibfnamefont {A.}~\bibnamefont {Parnachev}},
  \ and\ \bibinfo {author} {\bibfnamefont {G.}~\bibnamefont {Policastro}},\
  }\href {\doibase 10.1007/JHEP09(2014)010} {\bibfield  {journal} {\bibinfo
  {journal} {Journal of High Energy Physics}\ }\textbf {\bibinfo {volume}
  {2014}},\ \bibinfo {pages} {10} (\bibinfo {year} {2014})}\BibitemShut
  {NoStop}%
\bibitem [{\citenamefont {Calabrese}\ \emph {et~al.}(2015)\citenamefont
  {Calabrese}, \citenamefont {Cardy},\ and\ \citenamefont
  {Tonni}}]{calabrese2015}%
  \BibitemOpen
  \bibfield  {author} {\bibinfo {author} {\bibfnamefont {P.}~\bibnamefont
  {Calabrese}}, \bibinfo {author} {\bibfnamefont {J.}~\bibnamefont {Cardy}}, \
  and\ \bibinfo {author} {\bibfnamefont {E.}~\bibnamefont {Tonni}},\ }\href
  {http://stacks.iop.org/1751-8121/48/i=1/a=015006} {\bibfield  {journal}
  {\bibinfo  {journal} {Journal of Physics A: Mathematical and Theoretical}\
  }\textbf {\bibinfo {volume} {48}},\ \bibinfo {pages} {015006} (\bibinfo
  {year} {2015})}\BibitemShut {NoStop}%
\bibitem [{\citenamefont {Nobili}\ \emph {et~al.}(2015)\citenamefont {Nobili},
  \citenamefont {Coser},\ and\ \citenamefont
  {Tonni}}]{Tonni_negativity_cft_2015}%
  \BibitemOpen
  \bibfield  {author} {\bibinfo {author} {\bibfnamefont {C.~D.}\ \bibnamefont
  {Nobili}}, \bibinfo {author} {\bibfnamefont {A.}~\bibnamefont {Coser}}, \
  and\ \bibinfo {author} {\bibfnamefont {E.}~\bibnamefont {Tonni}},\ }\href
  {\doibase 10.1088/1742-5468/2015/06/p06021} {\bibfield  {journal} {\bibinfo
  {journal} {Journal of Statistical Mechanics: Theory and Experiment}\ }\textbf
  {\bibinfo {volume} {2015}},\ \bibinfo {pages} {P06021} (\bibinfo {year}
  {2015})}\BibitemShut {NoStop}%
\bibitem [{\citenamefont {Wichterich}\ \emph {et~al.}(2009)\citenamefont
  {Wichterich}, \citenamefont {Molina-Vilaplana},\ and\ \citenamefont
  {Bose}}]{Bose_2009_spin_chains}%
  \BibitemOpen
  \bibfield  {author} {\bibinfo {author} {\bibfnamefont {H.}~\bibnamefont
  {Wichterich}}, \bibinfo {author} {\bibfnamefont {J.}~\bibnamefont
  {Molina-Vilaplana}}, \ and\ \bibinfo {author} {\bibfnamefont
  {S.}~\bibnamefont {Bose}},\ }\href {\doibase 10.1103/PhysRevA.80.010304}
  {\bibfield  {journal} {\bibinfo  {journal} {Phys. Rev. A}\ }\textbf {\bibinfo
  {volume} {80}},\ \bibinfo {pages} {010304} (\bibinfo {year}
  {2009})}\BibitemShut {NoStop}%
\bibitem [{\citenamefont {Calabrese}\ \emph {et~al.}(2013)\citenamefont
  {Calabrese}, \citenamefont {Tagliacozzo},\ and\ \citenamefont
  {Tonni}}]{Calabrese_2013_critical_ising}%
  \BibitemOpen
  \bibfield  {author} {\bibinfo {author} {\bibfnamefont {P.}~\bibnamefont
  {Calabrese}}, \bibinfo {author} {\bibfnamefont {L.}~\bibnamefont
  {Tagliacozzo}}, \ and\ \bibinfo {author} {\bibfnamefont {E.}~\bibnamefont
  {Tonni}},\ }\href {\doibase 10.1088/1742-5468/2013/05/p05002} {\bibfield
  {journal} {\bibinfo  {journal} {Journal of Statistical Mechanics: Theory and
  Experiment}\ }\textbf {\bibinfo {volume} {2013}},\ \bibinfo {pages} {P05002}
  (\bibinfo {year} {2013})}\BibitemShut {NoStop}%
\bibitem [{\citenamefont {Ruggiero}\ \emph {et~al.}(2016)\citenamefont
  {Ruggiero}, \citenamefont {Alba},\ and\ \citenamefont
  {Calabrese}}]{random_spin_chain_2016}%
  \BibitemOpen
  \bibfield  {author} {\bibinfo {author} {\bibfnamefont {P.}~\bibnamefont
  {Ruggiero}}, \bibinfo {author} {\bibfnamefont {V.}~\bibnamefont {Alba}}, \
  and\ \bibinfo {author} {\bibfnamefont {P.}~\bibnamefont {Calabrese}},\ }\href
  {\doibase 10.1103/PhysRevB.94.035152} {\bibfield  {journal} {\bibinfo
  {journal} {Phys. Rev. B}\ }\textbf {\bibinfo {volume} {94}},\ \bibinfo
  {pages} {035152} (\bibinfo {year} {2016})}\BibitemShut {NoStop}%
\bibitem [{\citenamefont {Gray}(2018)}]{gray2018fast}%
  \BibitemOpen
  \bibfield  {author} {\bibinfo {author} {\bibfnamefont {J.}~\bibnamefont
  {Gray}},\ }\href@noop {} {\bibfield  {journal} {\bibinfo  {journal} {arXiv
  preprint arXiv:1809.01685}\ } (\bibinfo {year} {2018})}\BibitemShut {NoStop}%
\bibitem [{\citenamefont {Javanmard}\ \emph {et~al.}(2018)\citenamefont
  {Javanmard}, \citenamefont {Trapin}, \citenamefont {Bera}, \citenamefont
  {Bardarson},\ and\ \citenamefont {Heyl}}]{negativity_ising_chain_2018}%
  \BibitemOpen
  \bibfield  {author} {\bibinfo {author} {\bibfnamefont {Y.}~\bibnamefont
  {Javanmard}}, \bibinfo {author} {\bibfnamefont {D.}~\bibnamefont {Trapin}},
  \bibinfo {author} {\bibfnamefont {S.}~\bibnamefont {Bera}}, \bibinfo {author}
  {\bibfnamefont {J.~H.}\ \bibnamefont {Bardarson}}, \ and\ \bibinfo {author}
  {\bibfnamefont {M.}~\bibnamefont {Heyl}},\ }\href {\doibase
  10.1088/1367-2630/aad9ba} {\bibfield  {journal} {\bibinfo  {journal} {New
  Journal of Physics}\ }\textbf {\bibinfo {volume} {20}},\ \bibinfo {pages}
  {083032} (\bibinfo {year} {2018})}\BibitemShut {NoStop}%
\bibitem [{\citenamefont {Turkeshi}\ \emph {et~al.}(2020)\citenamefont
  {Turkeshi}, \citenamefont {Ruggiero},\ and\ \citenamefont
  {Calabrese}}]{negativity_random_singlet}%
  \BibitemOpen
  \bibfield  {author} {\bibinfo {author} {\bibfnamefont {X.}~\bibnamefont
  {Turkeshi}}, \bibinfo {author} {\bibfnamefont {P.}~\bibnamefont {Ruggiero}},
  \ and\ \bibinfo {author} {\bibfnamefont {P.}~\bibnamefont {Calabrese}},\
  }\href {\doibase 10.1103/PhysRevB.101.064207} {\bibfield  {journal} {\bibinfo
   {journal} {Phys. Rev. B}\ }\textbf {\bibinfo {volume} {101}},\ \bibinfo
  {pages} {064207} (\bibinfo {year} {2020})}\BibitemShut {NoStop}%
\bibitem [{\citenamefont {Lee}\ and\ \citenamefont {Vidal}(2013)}]{vidal2013}%
  \BibitemOpen
  \bibfield  {author} {\bibinfo {author} {\bibfnamefont {Y.~A.}\ \bibnamefont
  {Lee}}\ and\ \bibinfo {author} {\bibfnamefont {G.}~\bibnamefont {Vidal}},\
  }\href {\doibase 10.1103/PhysRevA.88.042318} {\bibfield  {journal} {\bibinfo
  {journal} {Phys. Rev. A}\ }\textbf {\bibinfo {volume} {88}},\ \bibinfo
  {pages} {042318} (\bibinfo {year} {2013})}\BibitemShut {NoStop}%
\bibitem [{\citenamefont {Castelnovo}(2013)}]{castelnovo2013}%
  \BibitemOpen
  \bibfield  {author} {\bibinfo {author} {\bibfnamefont {C.}~\bibnamefont
  {Castelnovo}},\ }\href {\doibase 10.1103/PhysRevA.88.042319} {\bibfield
  {journal} {\bibinfo  {journal} {Phys. Rev. A}\ }\textbf {\bibinfo {volume}
  {88}},\ \bibinfo {pages} {042319} (\bibinfo {year} {2013})}\BibitemShut
  {NoStop}%
\bibitem [{\citenamefont {Wen}\ \emph {et~al.}(2016{\natexlab{a}})\citenamefont
  {Wen}, \citenamefont {Chang},\ and\ \citenamefont
  {Ryu}}]{wen2016topological}%
  \BibitemOpen
  \bibfield  {author} {\bibinfo {author} {\bibfnamefont {X.}~\bibnamefont
  {Wen}}, \bibinfo {author} {\bibfnamefont {P.-Y.}\ \bibnamefont {Chang}}, \
  and\ \bibinfo {author} {\bibfnamefont {S.}~\bibnamefont {Ryu}},\ }\href
  {\doibase 10.1007/JHEP09(2016)012} {\bibfield  {journal} {\bibinfo  {journal}
  {Journal of High Energy Physics}\ }\textbf {\bibinfo {volume} {2016}},\
  \bibinfo {pages} {12} (\bibinfo {year} {2016}{\natexlab{a}})}\BibitemShut
  {NoStop}%
\bibitem [{\citenamefont {Wen}\ \emph {et~al.}(2016{\natexlab{b}})\citenamefont
  {Wen}, \citenamefont {Matsuura},\ and\ \citenamefont {Ryu}}]{wen2016edge}%
  \BibitemOpen
  \bibfield  {author} {\bibinfo {author} {\bibfnamefont {X.}~\bibnamefont
  {Wen}}, \bibinfo {author} {\bibfnamefont {S.}~\bibnamefont {Matsuura}}, \
  and\ \bibinfo {author} {\bibfnamefont {S.}~\bibnamefont {Ryu}},\ }\href
  {\doibase 10.1103/PhysRevB.93.245140} {\bibfield  {journal} {\bibinfo
  {journal} {Phys. Rev. B}\ }\textbf {\bibinfo {volume} {93}},\ \bibinfo
  {pages} {245140} (\bibinfo {year} {2016}{\natexlab{b}})}\BibitemShut
  {NoStop}%
\bibitem [{\citenamefont {Hart}\ and\ \citenamefont
  {Castelnovo}(2018)}]{castelnovo2018}%
  \BibitemOpen
  \bibfield  {author} {\bibinfo {author} {\bibfnamefont {O.}~\bibnamefont
  {Hart}}\ and\ \bibinfo {author} {\bibfnamefont {C.}~\bibnamefont
  {Castelnovo}},\ }\href {\doibase 10.1103/PhysRevB.97.144410} {\bibfield
  {journal} {\bibinfo  {journal} {Phys. Rev. B}\ }\textbf {\bibinfo {volume}
  {97}},\ \bibinfo {pages} {144410} (\bibinfo {year} {2018})}\BibitemShut
  {NoStop}%
\bibitem [{\citenamefont {Sang}\ \emph {et~al.}(2021)\citenamefont {Sang},
  \citenamefont {Li}, \citenamefont {Zhou}, \citenamefont {Chen}, \citenamefont
  {Hsieh},\ and\ \citenamefont {Fisher}}]{nega_hybrid_fisher}%
  \BibitemOpen
  \bibfield  {author} {\bibinfo {author} {\bibfnamefont {S.}~\bibnamefont
  {Sang}}, \bibinfo {author} {\bibfnamefont {Y.}~\bibnamefont {Li}}, \bibinfo
  {author} {\bibfnamefont {T.}~\bibnamefont {Zhou}}, \bibinfo {author}
  {\bibfnamefont {X.}~\bibnamefont {Chen}}, \bibinfo {author} {\bibfnamefont
  {T.~H.}\ \bibnamefont {Hsieh}}, \ and\ \bibinfo {author} {\bibfnamefont
  {M.~P.}\ \bibnamefont {Fisher}},\ }\href {\doibase
  10.1103/PRXQuantum.2.030313} {\bibfield  {journal} {\bibinfo  {journal} {PRX
  Quantum}\ }\textbf {\bibinfo {volume} {2}},\ \bibinfo {pages} {030313}
  (\bibinfo {year} {2021})}\BibitemShut {NoStop}%
\bibitem [{\citenamefont {Shi}\ \emph {et~al.}(2020)\citenamefont {Shi},
  \citenamefont {Dai},\ and\ \citenamefont {Lu}}]{nega_hybrid_shi}%
  \BibitemOpen
  \bibfield  {author} {\bibinfo {author} {\bibfnamefont {B.}~\bibnamefont
  {Shi}}, \bibinfo {author} {\bibfnamefont {X.}~\bibnamefont {Dai}}, \ and\
  \bibinfo {author} {\bibfnamefont {Y.-M.}\ \bibnamefont {Lu}},\ }\href@noop {}
  {\bibfield  {journal} {\bibinfo  {journal} {arXiv preprint arXiv:2012.00040}\
  } (\bibinfo {year} {2020})}\BibitemShut {NoStop}%
\bibitem [{\citenamefont {Levin}\ and\ \citenamefont
  {Wen}(2006)}]{levin2006detecting}%
  \BibitemOpen
  \bibfield  {author} {\bibinfo {author} {\bibfnamefont {M.}~\bibnamefont
  {Levin}}\ and\ \bibinfo {author} {\bibfnamefont {X.-G.}\ \bibnamefont
  {Wen}},\ }\href {\doibase 10.1103/PhysRevLett.96.110405} {\bibfield
  {journal} {\bibinfo  {journal} {Phys. Rev. Lett.}\ }\textbf {\bibinfo
  {volume} {96}},\ \bibinfo {pages} {110405} (\bibinfo {year}
  {2006})}\BibitemShut {NoStop}%
\bibitem [{\citenamefont {Kitaev}\ and\ \citenamefont
  {Preskill}(2006)}]{Kitaev06_1}%
  \BibitemOpen
  \bibfield  {author} {\bibinfo {author} {\bibfnamefont {A.}~\bibnamefont
  {Kitaev}}\ and\ \bibinfo {author} {\bibfnamefont {J.}~\bibnamefont
  {Preskill}},\ }\href {\doibase 10.1103/PhysRevLett.96.110404} {\bibfield
  {journal} {\bibinfo  {journal} {Phys. Rev. Lett.}\ }\textbf {\bibinfo
  {volume} {96}},\ \bibinfo {pages} {110404} (\bibinfo {year}
  {2006})}\BibitemShut {NoStop}%
\bibitem [{\citenamefont {Chen}\ \emph
  {et~al.}(2011{\natexlab{a}})\citenamefont {Chen}, \citenamefont {Gu},\ and\
  \citenamefont {Wen}}]{spt_1d_2011}%
  \BibitemOpen
  \bibfield  {author} {\bibinfo {author} {\bibfnamefont {X.}~\bibnamefont
  {Chen}}, \bibinfo {author} {\bibfnamefont {Z.-C.}\ \bibnamefont {Gu}}, \ and\
  \bibinfo {author} {\bibfnamefont {X.-G.}\ \bibnamefont {Wen}},\ }\href
  {\doibase 10.1103/PhysRevB.83.035107} {\bibfield  {journal} {\bibinfo
  {journal} {Phys. Rev. B}\ }\textbf {\bibinfo {volume} {83}},\ \bibinfo
  {pages} {035107} (\bibinfo {year} {2011}{\natexlab{a}})}\BibitemShut
  {NoStop}%
\bibitem [{\citenamefont {Chen}\ \emph
  {et~al.}(2011{\natexlab{b}})\citenamefont {Chen}, \citenamefont {Gu},\ and\
  \citenamefont {Wen}}]{spt_2011}%
  \BibitemOpen
  \bibfield  {author} {\bibinfo {author} {\bibfnamefont {X.}~\bibnamefont
  {Chen}}, \bibinfo {author} {\bibfnamefont {Z.-C.}\ \bibnamefont {Gu}}, \ and\
  \bibinfo {author} {\bibfnamefont {X.-G.}\ \bibnamefont {Wen}},\ }\href
  {\doibase 10.1103/PhysRevB.84.235128} {\bibfield  {journal} {\bibinfo
  {journal} {Phys. Rev. B}\ }\textbf {\bibinfo {volume} {84}},\ \bibinfo
  {pages} {235128} (\bibinfo {year} {2011}{\natexlab{b}})}\BibitemShut
  {NoStop}%
\bibitem [{\citenamefont {You}\ \emph {et~al.}(2014)\citenamefont {You},
  \citenamefont {Bi}, \citenamefont {Rasmussen}, \citenamefont {Slagle},\ and\
  \citenamefont {Xu}}]{You_strange_correlator_2014}%
  \BibitemOpen
  \bibfield  {author} {\bibinfo {author} {\bibfnamefont {Y.-Z.}\ \bibnamefont
  {You}}, \bibinfo {author} {\bibfnamefont {Z.}~\bibnamefont {Bi}}, \bibinfo
  {author} {\bibfnamefont {A.}~\bibnamefont {Rasmussen}}, \bibinfo {author}
  {\bibfnamefont {K.}~\bibnamefont {Slagle}}, \ and\ \bibinfo {author}
  {\bibfnamefont {C.}~\bibnamefont {Xu}},\ }\href {\doibase
  10.1103/PhysRevLett.112.247202} {\bibfield  {journal} {\bibinfo  {journal}
  {Phys. Rev. Lett.}\ }\textbf {\bibinfo {volume} {112}},\ \bibinfo {pages}
  {247202} (\bibinfo {year} {2014})}\BibitemShut {NoStop}%
\bibitem [{\citenamefont {Lu}\ and\ \citenamefont {Vijay}()}]{New_Paper}%
  \BibitemOpen
  \bibfield  {author} {\bibinfo {author} {\bibfnamefont {T.-C.}\ \bibnamefont
  {Lu}}\ and\ \bibinfo {author} {\bibfnamefont {S.}~\bibnamefont {Vijay}},\
  }\href@noop {} {\bibinfo  {journal} {Unpublished}\ }\BibitemShut {NoStop}%
\bibitem [{\citenamefont {Gaiotto}\ \emph {et~al.}(2015)\citenamefont
  {Gaiotto}, \citenamefont {Kapustin}, \citenamefont {Seiberg},\ and\
  \citenamefont {Willett}}]{higher_form_2015}%
  \BibitemOpen
\bibfield  {journal} {  }\bibfield  {author} {\bibinfo {author} {\bibfnamefont
  {D.}~\bibnamefont {Gaiotto}}, \bibinfo {author} {\bibfnamefont
  {A.}~\bibnamefont {Kapustin}}, \bibinfo {author} {\bibfnamefont
  {N.}~\bibnamefont {Seiberg}}, \ and\ \bibinfo {author} {\bibfnamefont
  {B.}~\bibnamefont {Willett}},\ }\href {\doibase 10.1007/JHEP02(2015)172}
  {\bibfield  {journal} {\bibinfo  {journal} {Journal of High Energy Physics}\
  }\textbf {\bibinfo {volume} {2015}},\ \bibinfo {pages} {172} (\bibinfo {year}
  {2015})}\BibitemShut {NoStop}%
\bibitem [{\citenamefont {Lu}\ and\ \citenamefont
  {Grover}(2020)}]{lu2019structure}%
  \BibitemOpen
  \bibfield  {author} {\bibinfo {author} {\bibfnamefont {T.-C.}\ \bibnamefont
  {Lu}}\ and\ \bibinfo {author} {\bibfnamefont {T.}~\bibnamefont {Grover}},\
  }\href {\doibase 10.1103/PhysRevResearch.2.043345} {\bibfield  {journal}
  {\bibinfo  {journal} {Phys. Rev. Research}\ }\textbf {\bibinfo {volume}
  {2}},\ \bibinfo {pages} {043345} (\bibinfo {year} {2020})}\BibitemShut
  {NoStop}%
\bibitem [{\citenamefont {Chen}\ \emph {et~al.}(2014)\citenamefont {Chen},
  \citenamefont {Lu},\ and\ \citenamefont {Vishwanath}}]{chen_decorated}%
  \BibitemOpen
  \bibfield  {author} {\bibinfo {author} {\bibfnamefont {X.}~\bibnamefont
  {Chen}}, \bibinfo {author} {\bibfnamefont {Y.-M.}\ \bibnamefont {Lu}}, \ and\
  \bibinfo {author} {\bibfnamefont {A.}~\bibnamefont {Vishwanath}},\ }\href
  {\doibase 10.1038/ncomms4507} {\bibfield  {journal} {\bibinfo  {journal}
  {Nature Communications}\ }\textbf {\bibinfo {volume} {5}},\ \bibinfo {pages}
  {3507} (\bibinfo {year} {2014})}\BibitemShut {NoStop}%
\bibitem [{app()}]{appendix}%
  \BibitemOpen
  \href@noop {} {\bibinfo  {journal} {See supplemental material}\ }\BibitemShut
  {NoStop}%
\bibitem [{\citenamefont {Yoshida}(2016)}]{yoshida_2016_symmetry}%
  \BibitemOpen
\bibfield  {journal} {  }\bibfield  {author} {\bibinfo {author} {\bibfnamefont
  {B.}~\bibnamefont {Yoshida}},\ }\href {\doibase 10.1103/PhysRevB.93.155131}
  {\bibfield  {journal} {\bibinfo  {journal} {Phys. Rev. B}\ }\textbf {\bibinfo
  {volume} {93}},\ \bibinfo {pages} {155131} (\bibinfo {year}
  {2016})}\BibitemShut {NoStop}%
\bibitem [{\citenamefont {Mazac}\ and\ \citenamefont
  {Hamma}(2012)}]{hamma2012topological}%
  \BibitemOpen
  \bibfield  {author} {\bibinfo {author} {\bibfnamefont {D.}~\bibnamefont
  {Mazac}}\ and\ \bibinfo {author} {\bibfnamefont {A.}~\bibnamefont {Hamma}},\
  }\href {\doibase https://doi.org/10.1016/j.aop.2012.05.004} {\bibfield
  {journal} {\bibinfo  {journal} {Annals of Physics}\ }\textbf {\bibinfo
  {volume} {327}},\ \bibinfo {pages} {2096 } (\bibinfo {year}
  {2012})}\BibitemShut {NoStop}%
\bibitem [{\citenamefont {Castelnovo}\ and\ \citenamefont
  {Chamon}(2008)}]{castelnovo2008topological}%
  \BibitemOpen
  \bibfield  {author} {\bibinfo {author} {\bibfnamefont {C.}~\bibnamefont
  {Castelnovo}}\ and\ \bibinfo {author} {\bibfnamefont {C.}~\bibnamefont
  {Chamon}},\ }\href {\doibase 10.1103/PhysRevB.78.155120} {\bibfield
  {journal} {\bibinfo  {journal} {Phys. Rev. B}\ }\textbf {\bibinfo {volume}
  {78}},\ \bibinfo {pages} {155120} (\bibinfo {year} {2008})}\BibitemShut
  {NoStop}%
\bibitem [{\citenamefont {Hastings}\ and\ \citenamefont
  {Wen}(2005)}]{hastings_2005_gauge}%
  \BibitemOpen
  \bibfield  {author} {\bibinfo {author} {\bibfnamefont {M.~B.}\ \bibnamefont
  {Hastings}}\ and\ \bibinfo {author} {\bibfnamefont {X.-G.}\ \bibnamefont
  {Wen}},\ }\href {\doibase 10.1103/PhysRevB.72.045141} {\bibfield  {journal}
  {\bibinfo  {journal} {Phys. Rev. B}\ }\textbf {\bibinfo {volume} {72}},\
  \bibinfo {pages} {045141} (\bibinfo {year} {2005})}\BibitemShut {NoStop}%
\bibitem [{\citenamefont {Baxter}(2016)}]{baxter2016exactly}%
  \BibitemOpen
  \bibfield  {author} {\bibinfo {author} {\bibfnamefont {R.~J.}\ \bibnamefont
  {Baxter}},\ }\href@noop {} {\emph {\bibinfo {title} {Exactly solved models in
  statistical mechanics}}}\ (\bibinfo  {publisher} {Elsevier},\ \bibinfo {year}
  {2016})\BibitemShut {NoStop}%
\bibitem [{\citenamefont {Raussendorf}\ \emph {et~al.}(2005)\citenamefont
  {Raussendorf}, \citenamefont {Bravyi},\ and\ \citenamefont
  {Harrington}}]{3d_cluster_state_2005}%
  \BibitemOpen
  \bibfield  {author} {\bibinfo {author} {\bibfnamefont {R.}~\bibnamefont
  {Raussendorf}}, \bibinfo {author} {\bibfnamefont {S.}~\bibnamefont {Bravyi}},
  \ and\ \bibinfo {author} {\bibfnamefont {J.}~\bibnamefont {Harrington}},\
  }\href {\doibase 10.1103/PhysRevA.71.062313} {\bibfield  {journal} {\bibinfo
  {journal} {Phys. Rev. A}\ }\textbf {\bibinfo {volume} {71}},\ \bibinfo
  {pages} {062313} (\bibinfo {year} {2005})}\BibitemShut {NoStop}%
\bibitem [{\citenamefont {Wang}\ \emph {et~al.}(2015)\citenamefont {Wang},
  \citenamefont {Nahum},\ and\ \citenamefont {Senthil}}]{Chong_2015_loop}%
  \BibitemOpen
  \bibfield  {author} {\bibinfo {author} {\bibfnamefont {C.}~\bibnamefont
  {Wang}}, \bibinfo {author} {\bibfnamefont {A.}~\bibnamefont {Nahum}}, \ and\
  \bibinfo {author} {\bibfnamefont {T.}~\bibnamefont {Senthil}},\ }\href
  {\doibase 10.1103/PhysRevB.91.195131} {\bibfield  {journal} {\bibinfo
  {journal} {Phys. Rev. B}\ }\textbf {\bibinfo {volume} {91}},\ \bibinfo
  {pages} {195131} (\bibinfo {year} {2015})}\BibitemShut {NoStop}%
\bibitem [{\citenamefont {Fradkin}\ and\ \citenamefont
  {Shenker}(1979)}]{fradkin_shenker_diagram}%
  \BibitemOpen
  \bibfield  {author} {\bibinfo {author} {\bibfnamefont {E.}~\bibnamefont
  {Fradkin}}\ and\ \bibinfo {author} {\bibfnamefont {S.~H.}\ \bibnamefont
  {Shenker}},\ }\href {\doibase 10.1103/PhysRevD.19.3682} {\bibfield  {journal}
  {\bibinfo  {journal} {Phys. Rev. D}\ }\textbf {\bibinfo {volume} {19}},\
  \bibinfo {pages} {3682} (\bibinfo {year} {1979})}\BibitemShut {NoStop}%
\bibitem [{\citenamefont {Jongeward}\ \emph {et~al.}(1980)\citenamefont
  {Jongeward}, \citenamefont {Stack},\ and\ \citenamefont
  {Jayaprakash}}]{gauge_mc_calculation_Jayaprakash}%
  \BibitemOpen
  \bibfield  {author} {\bibinfo {author} {\bibfnamefont {G.~A.}\ \bibnamefont
  {Jongeward}}, \bibinfo {author} {\bibfnamefont {J.~D.}\ \bibnamefont
  {Stack}}, \ and\ \bibinfo {author} {\bibfnamefont {C.}~\bibnamefont
  {Jayaprakash}},\ }\href {\doibase 10.1103/PhysRevD.21.3360} {\bibfield
  {journal} {\bibinfo  {journal} {Phys. Rev. D}\ }\textbf {\bibinfo {volume}
  {21}},\ \bibinfo {pages} {3360} (\bibinfo {year} {1980})}\BibitemShut
  {NoStop}%
\bibitem [{\citenamefont {Tupitsyn}\ \emph {et~al.}(2010)\citenamefont
  {Tupitsyn}, \citenamefont {Kitaev}, \citenamefont {Prokof'ev},\ and\
  \citenamefont {Stamp}}]{gauge_phase_diagram_Stamp}%
  \BibitemOpen
  \bibfield  {author} {\bibinfo {author} {\bibfnamefont {I.~S.}\ \bibnamefont
  {Tupitsyn}}, \bibinfo {author} {\bibfnamefont {A.}~\bibnamefont {Kitaev}},
  \bibinfo {author} {\bibfnamefont {N.~V.}\ \bibnamefont {Prokof'ev}}, \ and\
  \bibinfo {author} {\bibfnamefont {P.~C.~E.}\ \bibnamefont {Stamp}},\ }\href
  {\doibase 10.1103/PhysRevB.82.085114} {\bibfield  {journal} {\bibinfo
  {journal} {Phys. Rev. B}\ }\textbf {\bibinfo {volume} {82}},\ \bibinfo
  {pages} {085114} (\bibinfo {year} {2010})}\BibitemShut {NoStop}%
\bibitem [{\citenamefont {Vidal}\ \emph {et~al.}(2009)\citenamefont {Vidal},
  \citenamefont {Dusuel},\ and\ \citenamefont
  {Schmidt}}]{gauge_phase_diagram_vidal}%
  \BibitemOpen
  \bibfield  {author} {\bibinfo {author} {\bibfnamefont {J.}~\bibnamefont
  {Vidal}}, \bibinfo {author} {\bibfnamefont {S.}~\bibnamefont {Dusuel}}, \
  and\ \bibinfo {author} {\bibfnamefont {K.~P.}\ \bibnamefont {Schmidt}},\
  }\href {\doibase 10.1103/PhysRevB.79.033109} {\bibfield  {journal} {\bibinfo
  {journal} {Phys. Rev. B}\ }\textbf {\bibinfo {volume} {79}},\ \bibinfo
  {pages} {033109} (\bibinfo {year} {2009})}\BibitemShut {NoStop}%
\bibitem [{\citenamefont {Somoza}\ \emph {et~al.}(2021)\citenamefont {Somoza},
  \citenamefont {Serna},\ and\ \citenamefont {Nahum}}]{nahum_dual_2020}%
  \BibitemOpen
  \bibfield  {author} {\bibinfo {author} {\bibfnamefont {A.~M.}\ \bibnamefont
  {Somoza}}, \bibinfo {author} {\bibfnamefont {P.}~\bibnamefont {Serna}}, \
  and\ \bibinfo {author} {\bibfnamefont {A.}~\bibnamefont {Nahum}},\ }\href
  {\doibase 10.1103/PhysRevX.11.041008} {\bibfield  {journal} {\bibinfo
  {journal} {Phys. Rev. X}\ }\textbf {\bibinfo {volume} {11}},\ \bibinfo
  {pages} {041008} (\bibinfo {year} {2021})}\BibitemShut {NoStop}%
\bibitem [{\citenamefont {Wen}(2019)}]{Wen_higer_symmetries_2019}%
  \BibitemOpen
  \bibfield  {author} {\bibinfo {author} {\bibfnamefont {X.-G.}\ \bibnamefont
  {Wen}},\ }\href {\doibase 10.1103/PhysRevB.99.205139} {\bibfield  {journal}
  {\bibinfo  {journal} {Phys. Rev. B}\ }\textbf {\bibinfo {volume} {99}},\
  \bibinfo {pages} {205139} (\bibinfo {year} {2019})}\BibitemShut {NoStop}%
\bibitem [{\citenamefont {Iqbal}\ and\ \citenamefont
  {McGreevy}(2021)}]{McGreevy_one_form_2021}%
  \BibitemOpen
  \bibfield  {author} {\bibinfo {author} {\bibfnamefont {N.}~\bibnamefont
  {Iqbal}}\ and\ \bibinfo {author} {\bibfnamefont {J.}~\bibnamefont
  {McGreevy}},\ }\href@noop {} {\bibfield  {journal} {\bibinfo  {journal}
  {arXiv preprint arXiv:2106.12610}\ } (\bibinfo {year} {2021})}\BibitemShut
  {NoStop}%
\bibitem [{\citenamefont {Vijay}\ \emph {et~al.}(2016)\citenamefont {Vijay},
  \citenamefont {Haah},\ and\ \citenamefont {Fu}}]{xcube}%
  \BibitemOpen
  \bibfield  {author} {\bibinfo {author} {\bibfnamefont {S.}~\bibnamefont
  {Vijay}}, \bibinfo {author} {\bibfnamefont {J.}~\bibnamefont {Haah}}, \ and\
  \bibinfo {author} {\bibfnamefont {L.}~\bibnamefont {Fu}},\ }\href {\doibase
  10.1103/PhysRevB.94.235157} {\bibfield  {journal} {\bibinfo  {journal} {Phys.
  Rev. B}\ }\textbf {\bibinfo {volume} {94}},\ \bibinfo {pages} {235157}
  (\bibinfo {year} {2016})}\BibitemShut {NoStop}%
\bibitem [{\citenamefont {Haah}(2011)}]{Haah}%
  \BibitemOpen
  \bibfield  {author} {\bibinfo {author} {\bibfnamefont {J.}~\bibnamefont
  {Haah}},\ }\href {\doibase 10.1103/PhysRevA.83.042330} {\bibfield  {journal}
  {\bibinfo  {journal} {Phys. Rev. A}\ }\textbf {\bibinfo {volume} {83}},\
  \bibinfo {pages} {042330} (\bibinfo {year} {2011})}\BibitemShut {NoStop}%
\bibitem [{\citenamefont {You}\ \emph {et~al.}(2018)\citenamefont {You},
  \citenamefont {Devakul}, \citenamefont {Burnell},\ and\ \citenamefont
  {Sondhi}}]{you_2018_subsystem}%
  \BibitemOpen
  \bibfield  {author} {\bibinfo {author} {\bibfnamefont {Y.}~\bibnamefont
  {You}}, \bibinfo {author} {\bibfnamefont {T.}~\bibnamefont {Devakul}},
  \bibinfo {author} {\bibfnamefont {F.~J.}\ \bibnamefont {Burnell}}, \ and\
  \bibinfo {author} {\bibfnamefont {S.~L.}\ \bibnamefont {Sondhi}},\ }\href
  {\doibase 10.1103/PhysRevB.98.035112} {\bibfield  {journal} {\bibinfo
  {journal} {Phys. Rev. B}\ }\textbf {\bibinfo {volume} {98}},\ \bibinfo
  {pages} {035112} (\bibinfo {year} {2018})}\BibitemShut {NoStop}%
\bibitem [{\citenamefont {Devakul}\ \emph {et~al.}(2019)\citenamefont
  {Devakul}, \citenamefont {You}, \citenamefont {Burnell},\ and\ \citenamefont
  {Sondhi}}]{fractal_spt_2019_sondhi}%
  \BibitemOpen
  \bibfield  {author} {\bibinfo {author} {\bibfnamefont {T.}~\bibnamefont
  {Devakul}}, \bibinfo {author} {\bibfnamefont {Y.}~\bibnamefont {You}},
  \bibinfo {author} {\bibfnamefont {F.~J.}\ \bibnamefont {Burnell}}, \ and\
  \bibinfo {author} {\bibfnamefont {S.~L.}\ \bibnamefont {Sondhi}},\ }\href
  {\doibase 10.21468/SciPostPhys.6.1.007} {\bibfield  {journal} {\bibinfo
  {journal} {SciPost Phys.}\ }\textbf {\bibinfo {volume} {6}},\ \bibinfo
  {pages} {7} (\bibinfo {year} {2019})}\BibitemShut {NoStop}%
\end{thebibliography}%
\renewcommand\refname{Reference}

\appendix

\onecolumngrid

\section*{ {  \large Supplemental Material } }

In this supplemental material, we provide details on statements of the main text. Appendix.\ref{appendix:general} presents the derivation of negativity spectrum for Gibbs states of stabilizer models. Appendix.\ref{appendix:parent} presents the derivation of the parent Hamiltonian for the state localized on the bipartition surface induced by partial transpose. Appendix.\ref{appendix:3d} and Appendix.\ref{appendix:4d} discuss the details on the 3d toric code and 4d toric code respectively.

\section{Negativity spectrum for  Gibbs states of stabilizer models}\label{appendix:general}

\subsection{General formalism}
Here we present an general formalism for taking a partial transpose for Gibbs states of stabilizer Hamiltonian $\hat{H}= -J \sum_{j=1}^N \hat{\theta}_j$, where each stabilizer $ \hat{\theta}_j$ is a tensor product of Pauli operators over lattice sites. The corresponding Gibbs state is $\hat{\rho} \sim  e^{-\beta  \hat{H} }  = e^{\beta J \sum_j  \hat{\theta}_j  }$, where $\sim$ is used to indicate that we omit the normalization. Utilizing the expansion $e^{\beta J \hat{\theta}_j }= \cosh(\beta J)+ \hat{\theta}_j \sinh(\beta J)= \cosh( \beta J )  \sum_{s_j=0,1} [\hat{\theta}_j\tanh(\beta J )  ]^{s_j } $, the Gibbs state can be written as a sum over stabilizer strings 
\begin{equation}
	\hat{\rho} \sim  \sum_{ \{ s_j  \}}  \left[ \prod_j  \hat{\theta}_j^{s_j }  \right] [\tanh(\beta J)]^{\sum_j s_j}, 
\end{equation}
where $s_j=0,1$ is a classical variable denoting the absence or presence of the stabilizer $\hat{\theta}_j$. By dividing the system into $\mA$ and its complement $\mB$ and taking a partial transpose over the region $\mA$ for stabilizer strings generates a sign  $ \phi(\{ s_j \}  )$
\begin{equation}
	\left[   \prod_j   \hat{\theta}_j^{s_j } \right]^{T_{\mA}} = \left[   \prod_j   \left[(\hat{\theta}_j)^{T_\mA}\right]^{s_j } \right] \phi(\{ s_j \}  ).
\end{equation}
where $  \phi (   \{ s_j  \}  ) = 1, -1 $. To determine the sign, we introduce  $\hat{\theta}_i  |_\mA$ to denote the part in $\hat{\theta}_i$ that acts non-trivially on $\mA$, and introduce a matrix $C$ that encodes the commutation relation between the restricted stabilizers, namely,  $C_{ij} =0, 1 $ for $[ \hat{\theta}_i|_\mA , \hat{\theta}_j |_\mA] =0$ and  $\{ \hat{\theta}_i|_\mA , \hat{\theta}_j |_\mA\} =0$. It follows that the sign resulting from partial transpose is 
\begin{equation}\label{appendix:sign}
	\phi(  \{ s_j   \} ) = (-1)^{\frac{1}{2 }s^{T}Cs    }  = (-1)^{\frac{1}{2 } \sum_{ij}  s_iC_{ij}s_j    } = (-1)^{  \sum_{i<j } s_i C_{ij}s_j  },
\end{equation}
which can take the value $1$ or $-1$ depending on whether the number of pairs of restricted stabilizer $\hat{\theta}_i|_A$ that anticommute is even or  odd. Therefore the partially transposed matrix reads 
\begin{equation} \label{appendix:eq1}
	\hat{\rho}^{T_\mA } \sim   \sum_{ \{ s_j  \}}  \left[ \prod_j  [( \hat{\theta}_j)^{T_\mA}]^{s_j }  \right]   [\tanh(\beta J)]^{\sum_j s_j}	\phi(  \{ s_j   \} ).
\end{equation}
Since $(\hat{\theta}_j)^{T_\mA}$ are mutually commuting, the negativity spectrum $\rho^{T_\mA}$, i.e. the eigenspectrum of $\hat{\rho}^{T_\mA}$, can be obtained by replacing $(\hat{\theta}_j)^{T_\mA}$ with number $ \theta_j  \in \{1, -1\}$ :

\begin{equation}
	\rho^{T_\mA } \sim   \sum_{ \{ s_j  \}}  \left[ \prod_j \theta_j^{s_j }  \right]   	\psi(\{ s_j \})  ,
\end{equation} 
where  
\begin{equation}\label{appendix:wave}
	\psi(\{ s_j \})= [\tanh(\beta J)]^{\sum_j s_j}  \phi( \{s_j   \}   ) =  [\tanh(\beta J)]^{\sum_j s_j}   (-1)^{\frac{1}{2 } \sum_{ij}  s_iC_{ij}s_j    }.  
\end{equation}
The negativity spectrum can be expressed as a strange correlator. To see this, we introduce a computational basis $\ket{\{s_j\}}$ with $ Z_j\ket{s_j } = (1-2 s_j) \ket{s_j}$ and the state $\ket{\psi}$: 
\begin{equation}
	\ket{ \psi   }  = \frac{1}{ \sqrt{ Z_0  } }  \sum_{\{s_j\}} 	\psi(  \{ s_j   \} )   \ket{\{ s_j  \}}    =  \frac{1}{\sqrt{Z_0}} \left[ \prod_i  \tanh(\beta J)^{  \frac{1- Z_i}{2} }  \right] \prod_{i< j } [CZ_{ij}]^{C_{ij}}  \ket{ + }. 
\end{equation}
The normalization constant is $Z_0 = \sum_{\{s_i \}}  [\psi(\{s_i\}) ]^2$. $\ket{+}$ is the $+1$ eigenstates of Pauli-X operators, and $CZ_{ij}$ is the two-qubit control-Z gate with the operation $CZ_{ij }\ket{s_i,s_j }=  (-1)^{  s_i s_j   }  \ket{s_i,s_j} $. In fact, one can derive the parent Hamiltonian of the state $\ket{\psi } $ (see Appendix.\ref{appendix:parent} for derivation): 

\begin{equation}
	H=  -  \sum_{j}  \left[ X_j \prod_{ \{i | C_{ij}=1    \} }  Z_i   - e^{   -  \alpha  Z_j  } \right],
\end{equation}
where  $\alpha=   -  \log \left[   \tanh(\beta J)   \right]$, and  the term $\prod_{ \{i | C_{ij}=1    \} }  Z_i $ denotes the product of $Z_i$ so that $\{  \theta_i|_{\mA} ,  \theta_j|_{\mA}    \}=0 $. Using the state $\ket{ \psi}$, the negativity spectrum $\rho^{T_\mA}$ in Eq.\ref{appendix:eq1} can be expressed as a strange correlator
\begin{equation}\label{appendix:strange}
	\rho^{T_{\mA }  }  =  \frac{1}{\mathcal{N}} \bra{+} \prod_j Z_j^{\frac{1- \theta_j  }{2}} \ket{ \psi},
\end{equation}
where we have used $\theta_j^{s_j} =  (1-2 s_j )^{ \frac{1-\theta_j}{2}  }   $, and  $\mathcal{N}$ is a normalization constant such that sum of all eigenvalues is one.

\begin{figure}
	\centering
	\begin{subfigure}[b]{0.5\textwidth}
		\includegraphics[width=\textwidth]{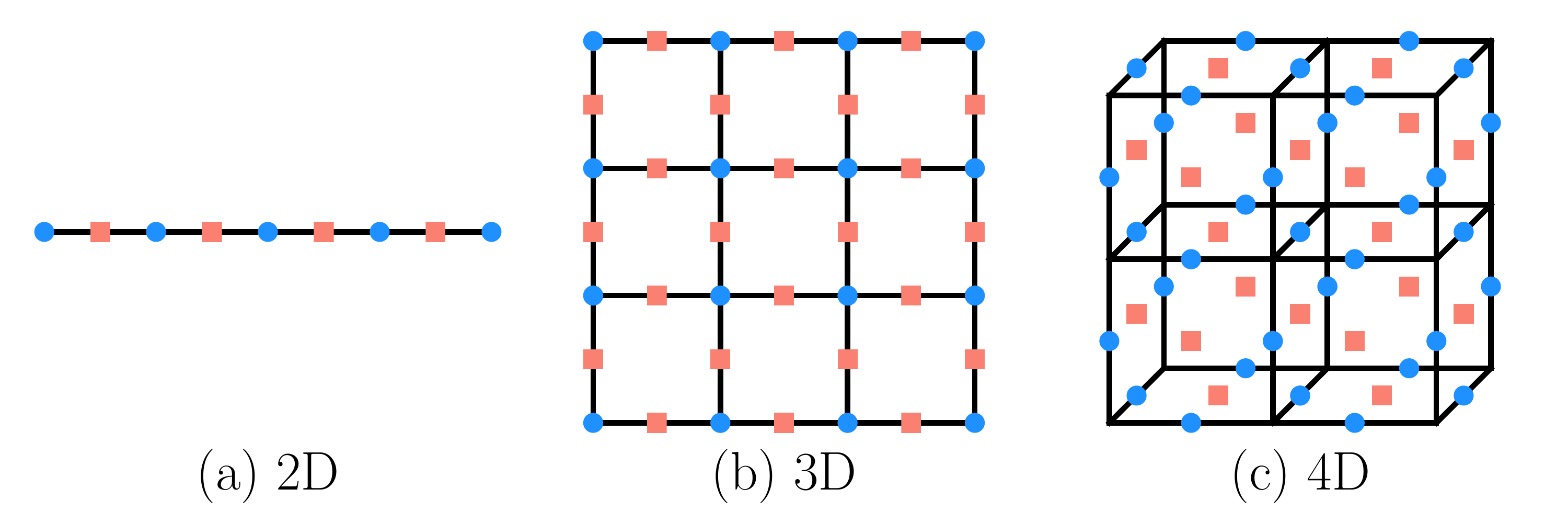}
	\end{subfigure}
	\caption{The location of boundary stabilizers in d-dim toric code, where blue circles and red squares label the lattice sites corresponding to $A_i$ and $B_j$ stabilizers. (a) 1d bipartition boundary in 2d toric code. (b) 2d bipartition boundary in 3d toric code. (c) 3d bipartition boundary in 4d toric code.}  \label{fig:boundary_stabilizers}
\end{figure}

Now we apply the  above formalism to the toric code Hamiltonian: $\hat{H}= -\lambda_A \sum_{i } \hat{A}_i -\lambda_B \sum_{j} \hat{B}_j $, where $\hat{A}_i$ and $\hat{B}_j$ denote the stabilizers consisting of Pauli-X and Pauli-Z operators respectively. First we divide the entire system into two subsystems $\mA$ and $\mB$, and write the Hamiltonian as $\hat{H}= \hat{H}_{\mA } + \hat{H}_{\mB }+ \hat{H}_{\partial }  $, where $\hat{H}_{\mA} (\hat{H}_{\mB})$ denotes the terms in $\hat{H}$ suppported on $\mA (\mB)$, and $\hat{H}_{\partial} $ denotes the interaction betwene $\mA$ and $\mB$. Since only the stabilizers acting on the bipartition boundary can anticommute when restricted in a subregion, the partial transpose  acts on the boundary part of $\hat{\rho}$ as $\hat{\rho}^{T_{\mA}}  \sim e^{ -\beta (\hat{H}_{\mA } + \hat{H}_{\mB}  ) }  \left( e^{ -\beta \hat{H}_{ \partial }   }  \right)^{T_{\mA}}$, and the non-trivial feature of spectrum of $\hat{\rho}^{T_{\mA}}$ solely comes from the boundary part $ \left( e^{ -\beta  \hat{H}_{\partial }   }  \right)^{T_{\mA}}$, which we derive below. Let $R_a$ and $R_b$ label the collection of lattice sites corresponding to the location of $A_i$ and $B_j$ stabilizers acting on the boundary (e.g. see Fig.\ref{fig:boundary_stabilizers}), applying Eq.\ref{appendix:strange} gives the following spectrum

\begin{equation}\label{appendix:toric}
	\left( e^{-\beta H_{\partial }}	   \right)^{T_\mA} \sim \bra{+}   \prod_{i \in R_a} Z_i^{ \frac{1-A_i}{2}}  \prod_{j \in R_b} Z_j^{ \frac{1-B_j}{2}}  \ket{ \psi}.
\end{equation}
where the state $\ket{\psi}$ lives in the Hilbert space spanned by $\ket{\{  a_i, b_j\}}$, and can be written as (via Eq.\ref{appendix:wave})

\begin{equation}\label{appendix:toric_sign}
	\ket{\psi}   =\frac{1}{ \sqrt{Z_0}}  \sum_{\{a_i,b_j \}}  [\tanh(\beta \lambda_A )]^{\sum_{i \in R_a} } [\tanh(\beta \lambda_B )]^{\sum_{j \in R_b} }   \phi(  \{a_i, b_j\} ), 
\end{equation}
where the sign $  \phi(  \{a_i, b_j\} ) =  \prod_{i \in R_a}(-1)^{a_i \sum_{ j\in \partial i  }b_j    }$ (via Eq.\ref{appendix:sign}) with $\sum_{j \in \partial i}$ indicating a summation over the sites $j\in R_b $ that are adjacent to the site $i \in R_a$. This is because any two adjacent boundary stabilizer $\hat{A}_i$ and $\hat{B}_j$ must anticommute when restricted on a subregion. Alternatively, the sign  $ \prod_{i \in R_a}(-1)^{a_i \sum_{ j\in \partial i  }b_j    } $  can be written as $ \prod_{j \in R_b}(-1)^{b_j \sum_{ i\in  \partial j  } a_i   }$. Therefore, Eq.\ref{appendix:toric} and Eq.\ref{appendix:toric_sign} show that the eigenspectrum of $	\left( e^{-\beta H_{\partial }}	   \right)^{T_\mA} $ are given by various choices of $\{A_i,B_j\}$, which correspond to various choices of Pauli-Z's insertion in the strange correlators. \\

\noindent\underline{Strange correlators as conventional correlators in classical statistical models:}\\

By explicitly computing the strange correlators, one finds that they can be expressed as multi-spin correlation functions of $\tau_i$ spins ($\tau_i = 1-2a_i =\pm 1$) in a classical statistical model: 
\begin{equation}\label{eq:spectrum_A}
	\left( e^{-\beta H_{ \partial }}	   \right)^{T_\mA}\sim  \sum_{   \{ \tau_{i} \} }  \left[\prod_i   {\tau _i}^{\frac{1-A_i   }{2}  }  \right] e^{ -   \tilde{H}_A  } ,
\end{equation}
where $\tilde{H}_A$  is consisting of the onsite-field terms as well as the interactions with a coupling strength specified by $B_j$: $\tilde{H}_A   = K_A  \sum_i  \frac{1-\tau_i }{2}  - \beta \lambda_B  \sum_{j } B_j   \prod_{i \in \partial j } \tau_i$ with $K_A= -   \log(\tanh(\beta \lambda_A)) $. Alternatively, the negativity spectrum can also be expressed as  multi-spin correlation functions of $\sigma_j =1-2b_j  =\pm1 $ in the corresponding ``dual'' classical model: 
\begin{equation}\label{eq:spectrum_B}
	\left( e^{-\beta H_{\partial }}	   \right)^{T_\mA}\sim  \sum_{   \{\sigma_{j} \} }  \left[\prod_j {\sigma_i}^{\frac{1-B_j  }{2}  }  \right] e^{ -   \tilde{H}_B }, 
\end{equation}
where  $	\tilde{H}_B  =  K_B  \sum_j  \frac{1- \sigma_j }{2}  -  \beta  \lambda_A  \sum_{i } A_i   \prod_{j \in \partial i }  \sigma_j$ with $K_B = -   \log(\tanh(\beta \lambda_B))$.  This  formalism allows us to  derive the statistical models that determines  the negativity spectrum for $d$-dim toric code.

\subsection{2d toric code}
The boundary of the 2d toric code involves alternating $A_1, B_1, A_2, B_2\cdots A_L, B_L$ stabilizers, and therefore one can define a 1d lattice with $A_i$ defined on the $i$-th site and $B_i$ defined on the link between the $i$ and $i$$+1$-sites. It follows that the  classical model describing the negativity spectrum is given by the 1d classical Ising model: $\tilde{H}_A=K_A  \sum_{i=1}^L  \frac{1-\tau_i}{2}   - \beta \lambda_B \sum_{i=1}^L B_i \tau_i \tau_{i+1}$. Alternatively, one can consider the dual description by $\tilde{H}_B$, which is again a 1d Ising model.

\subsection{3d toric code}
The boundary of the 3d toric code involves $A_i$ on lattices and $B_{ij}$ on links in a 2d lattice. The effective classical model describing the negativity spectrum is given by  a 2d classical Ising model: $\tilde{H}_A=K_A  \sum_{i}  \frac{1-\tau_i}{2}   - \beta \lambda_B \sum_{  \expval{ij} } B_{ij} \tau_i \tau_{j}$. Alternatively, one can consider the dual description by $\tilde{H}_B$, which is a 2d Ising gauge theory: $\tilde{H}_B = K_B \sum_{\expval{ij}} \frac{1-\sigma_{ij}}{2}   -\beta \lambda_A \sum_{i} A_i \prod_{ \expval{ij}|  i \in \partial \expval{ij}  } \sigma_{ij}$, where $\prod_{ \expval{ij}|  i \in \partial \expval{ij}  } \sigma_{ij}$ is the interaction between four $\sigma$ spins on links that share the same bounday site $i$.

\subsection{4d toric code}
The boundary of the 4d toric code involves $A_l$ on links and $B_{p}$ on plaquettes in a 3d lattice. The effective classical model describing the negativity spectrum is  a 3d classical Ising gauge theory: $\tilde{H}_A=K_A  \sum_{l}  \frac{1-\tau_l}{2}   - \beta \lambda_B \sum_{ p } B_{p} \prod_{l \in \partial p}\tau_l   $. Alternatively, one can consider the dual description by $\tilde{H}_B$, which is again a 3d classical Ising  gauge theory (but defined on the dual lattice): $\tilde{H}_B = K_B \sum_{ p  } \frac{1-\sigma_{p}}{2}   -\beta \lambda_A \sum_{l} A_l \prod_{p|  l \in \partial p  } \sigma_{p}$, where $\prod_{p|  l \in \partial p  } \sigma_{p}$ is the interaction between four $\sigma$ spins on plaquettes that share the same bounday link $l$.

\section{Derivation for the parent Hamiltonian of the state $\ket{\psi}$}\label{appendix:parent}

As discussed above, the negativity spectrum is $\rho^{T_{\mA }  }  =  \frac{1}{\mathcal{N}} \bra{+} \prod_j Z_j^{\frac{1- \theta_j  }{2}} \ket{ \psi}$, where  $	\ket{ \psi   }  \sim \left[ \prod_i  \tanh(\beta J)^{  \frac{1- Z_i}{2} }  \right] \prod_{i< j } [CZ_{ij}]^{C_{ij}}  \ket{ + }$. Here we present the derivation for the parent Hamiltonian of which $\ket{\psi}$ is the ground state. To start, we define $U_{cz}= \prod_{i< j } [CZ_{ij}]^{C_{ij}} $ and the operator

\begin{equation}
	Q_j= - U_{cz} X_j  U_{cz}^{\dagger}  +  e^{  -  \alpha Z_j}.
\end{equation}
with $\alpha=   -  \log \left[   \tanh(\beta J)   \right]$. A simple calculation shows that 
\begin{equation}
	Q_j  \ket{\psi}   \sim \left[ U_{cz} X_j  U_{cz}^{\dagger}  -  e^{  -  \alpha Z_j} \right]   e^{   \frac{\alpha}{2}  \sum_j Z_j    } U_{cz} \ket{+} =0. 
\end{equation}
On the other hand, $Q_j$ is a positive semi-definite matrix by noticing that it can be unitarily transformed to the matrix $-X_j + e^{ -  \alpha Z_j}$, which has non-negative  eigenvalues $0$ and $2\cosh(\alpha)$. Therefore, $\ket{\psi}$ is the exact ground state with zero energy of the Hamiltonian 
\begin{equation}
	H= \sum_j Q_j =  -  \sum_j \left[  U_{cz} X_j  U_{cz}^{\dagger}  -  e^{  -  \alpha Z_j}  \right]. 
\end{equation}

\section{Structure of negativity spectrum and  entanglement negativity  in 3d toric code}\label{appendix:3d}
\subsection{Structure of negativity spectrum}

The 3d toric code exhibits a topological order below a certain critical temperature $T_c$ when forbidding the point-like excitations. As a simplification, here we consider only the bipartition-boundary part of the density matrix by forbidding any excitations in the bulk. We show that the negativity spectrum encodes long-range braiding between two types of charges below $T_c$, and we derive the exact result of entanglement negativity at all temperatures.

When forbidding the bulk excitations, the negativity spectrum $\rho^{T_\mA}$ is solely given by $\left( e^{-\beta H_{\partial}} \right)^{T_\mA}$, which can be written as correlation functions in the 2d Ising model under a symmetry-breaking field (via Eq.\ref{eq:spectrum_A}):

\begin{equation}
	\rho^{T_\mA} ( \{A_i,B_{ij }\} )  \sim \bra{+}  \prod_i Z_i^{\frac{1-A_i }{2}} \prod_{\expval{ij}} Z_{ij}^{\frac{1-B_{ij}}{2}}   \ket{ \psi} \sim   \sum_{   \{ \tau_{i} \} }  \left[\prod_i   {\tau _i}^{\frac{1-A_i   }{2}  }  \right] e^{ -K_A \sum_i \frac{1-\tau_i}{2}  + \beta \lambda_B \sum_{\expval{ij}} B_{ij}\tau_i \tau_j      }.
\end{equation}
Here the constraint $\prod_{\expval{ij} \in  \partial p   } B_{ij }=1$ is imposed on every  plaquette $p$, which is a consequence of  the local constraint in the 3d toric code that the product of six $B_p$ stabilizers on the boundary of each cube equals identity as the bulk excitations are prohibited. $\{A_j\}$, $\{B_{ij}\}$ determine the choice for correlators and the sign of interactions between neighboring spins. The above expression suggests that a finite-temperature order can exist only when the symmetry-breaking field  $K_A =0$ (i.e. $\beta  \lambda_A \to \infty$), corresponding to prohibiting point-like excitations in Gibbs states at any temperatures. In this limit, the negativity spectrum is 

\begin{equation}
	\rho^{T_\mA} ( \{A_i,B_{ij }\} ) \sim \sum_{   \{ \tau_{i} \} }  \left[   \prod_i   {\tau _i}^{\frac{1-A_i   }{2}  }  \right] e^{ \beta \lambda_B \sum_{\expval{ij}} B_{ij}\tau_i \tau_j      }. 
\end{equation}
Due to the Ising symmetry in the Boltzmann weights, non-vanishing negativity spectrum requires the quantity $ \prod_i   {\tau _i}^{\frac{1-A_i   }{2}  }$ having even number of $\tau_i$ spins, which amounts to the constraint that $\prod_i A_i =1$.  Now let's analyze the sign structure of negativity spectrum. First consider the case with no charges, i.e. $A_i, B_{ij}=1$, the corresponding eigenvalue $\rho^{T_{\mA}}  \sim \sum_{\{\tau_i\}  }  e^{ \beta \lambda_B \sum_{\expval{ij}}\tau_i \tau_j      }$ is surely positive. Next, we consider $B_{ij}=1 \forall \expval{ij}$, which gives the eigenvalue $ \rho^{T_{\mA}}  \sim  \sum_{\{\tau_i\}}  \prod_{i} \tau_i^{\frac{1-A_i }{2}  } \sum_{\{\tau_i\}  }  e^{ \beta \lambda_B \sum_{\expval{ij}}\tau_i \tau_j      } $.  Choosing $A_i=A_j=-1$ gives the two-point correlation  $\sum_{\{\tau_i\}}  \tau_i \tau_j   e^{ \beta \lambda_B \sum_{\expval{ij}}\tau_i \tau_j  } $. This is again positive since it can be written as $\sum_{\{\tau_i\}}  \tau_i \tau_j    \prod_{\expval{ij}}   \left[  \cosh(\beta \lambda_B)  +   \tau_i \tau_j \sinh(\beta \lambda_B) \right] $, where one can expand the product $\prod_{\expval{ij}}$ and notice that only terms without containing $\tau_i$ spin variables will survive after the summation $\sum_{\{\tau_i \}  }$. Such an argument applies to the expectation value of $2n$-point functions for any integer $n$. We now consider a case with negative eigenvalue by flipping $\{A_i\}$ and $\{B_{ij}\}$ at the same time. Notice that the constraint $\prod_{\expval{ij}  \in  \partial p}B_{ij} =1 $ implies  different allowed $\{B_{ij}\}$ configurations are generated by flipping $B_{ij}$ on four links emanating from the site $i$. To have a negative eigenvalue of $\rho^{T_{\mA }}$, one can set $A_i =A_j =-1$ and  $B_{ij}=-1 $ on four links emanating from the site $i$. The corresponding eigenvalue is negative as can be seen by  making a local spin flip at the site $i$, giving rise to $	\rho^{T_\mA}  \sim   - \sum_{   \{ \tau_{i} \} }  \tau_i\tau_j e^{ \beta \lambda_B \sum_{\expval{ij}}\tau_i \tau_j      }$. Pictorially, one can connect the two lattice sites $i, j$ with $A_i=A_j=-1$ with the $A$-string, and construct a $B$-loop corresponding to four links with $B_{ij}=-1$. A minus sign results from the $A$-string piercing through the $B$-loop.

\subsection{Exact entanglement negativity}\label{appendix:3d_exact}
Utilizing the analysis of negativity spectrum above, we here derive the entanglement negativity for 3d toric code when point-like charges forbidden (same limit as considered above), and show that the transition of the topological order at finite temperature can be understood as a spontaneous symmetry breaking transition of the 2d Ising model. To start with, we utilize the negativity spectrum to write down the one-norm of the partially transposed Gibbs state:  $\norm{ \hat{\rho}^{T_\mA}}_1= \frac{Z_n}{Z_d}$ with 

\begin{equation}
	\begin{split}
		&Z_n= \sum_{\{A_i   \}}   \sum'_{ \{B_{ij}\}}  \abs{ \sum_{  \{\tau_i   \}  }\prod_{ i} \tau_i^{\frac{1-A_i}{2}}   e^{ \beta \lambda_B \sum_{\expval{ij}}  B_{ij} \tau_i \tau_j }} \\
		&Z_d=   \sum_{\{A_i   \} } \sum'_{ \{B_{ij}\}}  \sum_{  \{\tau_{i}   \}  }\prod_{ i} \tau_i^{\frac{1-A_i}{2}}   e^{ \beta \lambda_B \sum_{\expval{ij}}  B_{ij} \tau_i \tau_j },  
	\end{split}
\end{equation}
where $\sum_{\{B_{ij}\}}^{'}$ denotes a summation over $\{B_{ij}\}$ subject to the local constraint  $\prod_{\expval{ij} \in  \partial p   } B_{ij }=1$. For the denominator $Z_d$, it is straightforward to sum over $\{\tau_i \}$ and $\{A_i\}$ to find $Z_d= 2^{L^2} \sum'_{\{ B_{ij}\} }  e^{ \beta    \lambda_B  \sum_{\expval{ij}} B_{ij}}$. For the numerator $Z_n$,  using the fact that $\prod_{\expval{ij} \in \partial p} B_{ij}=1$ and the local gauge invariance of the Gibbs weight, namely, $\tau_i \to -\tau_i$ and  $B_{ij } \to -B_{ij}$ on four links emanating from the site $i$, one can remove  $B_{ij}$ in the Gibbs weight and find

\begin{equation}
	Z_n=  \sum_{\{A_i   \}}   \sum'_{ \{B_{ij}\}}  \abs{ \sum_{  \{\tau_i   \}  }\prod_{ i} \tau_i^{\frac{1-A_i}{2}}   e^{ \beta \lambda_B \sum_{\expval{ij}}  B_{ij} \tau_i \tau_j }}  =   \sum_{\{A_i   \}}   \sum'_{ \{B_{ij}\}}  \sum_{  \{\tau_i   \}  }\prod_{ i} \tau_i^{\frac{1-A_i}{2}}   e^{ \beta \lambda_B \sum_{\expval{ij}} \tau_i \tau_j } = 2^{L^2}     \sum'_{ \{B_{ij}\}} e^{2 \beta \lambda_B  L^2}  
\end{equation}
As a result, 
\begin{equation}\label{appendix:3d_toric_exact}
	\norm{\hat{\rho}^{T_\mA}}_1 = \frac{ \sum'_{\{   B_{ij} \} }  e^{2L^2 \beta 	\lambda_B}    }{    \sum'_{\{ B_{ij}\} }  e^{ \beta    \lambda_B  \sum_{\expval{ij}} B_{ij}}    }.
\end{equation}
One can introduce the dual variable $\tau_i$ on each lattice site by defining $B_{ij} = \tau_i \tau_j$ for neighboring sites so that the local constraint in $B_{ij}$ is implictly satisfied. It follows that

\begin{equation}
	\norm{\hat{\rho}^{T_\mA}}_1  = \frac{ \sum_{\{  \tau_{ i }    \} }  e^{2L^2 \beta 
			\lambda_B}    }{    \sum_{\{ \tau_l\} }  e^{ \beta    \lambda_B  \sum_{\expval{ij} \tau_i\tau_j  }  }    }  = \frac{2^{L^2}  e^{ -\beta E_g}    }{Z(T)}.
\end{equation}
where $E_g =-2 \lambda_BL^2  $ and $Z(T)=   \sum_{\{ \tau_l\} }  e^{ \beta    \lambda_B  \sum_{\expval{ij}} \tau_i\tau_j    }   $ is the energy and partition function for the 2d Ising model. Taking a logarithm gives the negativity 
\begin{equation}
	E_N=  L^2 \log 2  -\beta E_g        - \log Z(T)  
\end{equation}
Such an expression shows that in the 3d toric code where the bulk excitations and all point-like excitations forbidden, the negativity relates to the free energy of the 2d classical Ising model, which therefore exhibits a singularity across a finite critical temperature $T_c$. In particular, one can extract the topological negativity, i.e. the long-range component of negativity, by canceling out the short-range area-law component of negativity: $E_{\textrm{topo}}=  E_N(2N)-2E_N(N)=   -\log Z(2N,T)+ 2 \log Z(N,T)   $ with $N=L^2$ being the number of 2d lattice sites. In the thermodynamic limit, one expects $Z(N,T>T_c)= e^{ -\beta Nf(T)}$ for $T>T_c$, where $f$ is the free energy density. In contrast, for $T<T_c$ (the ordered phase), the existence of two spontaneous symmetry breaking sectors implies a universal prefactor in the partition function $Z(N,T)= 2e^{-\beta Nf(T)}$. Consequently, the topological part of the negativity $E_{\textrm{topo}}$ exhibits a discontinuity at $T_c$: 
\begin{equation}
	E_{\textrm{topo}} = \begin{cases} 
		\log 2 \quad  \text{for } T < T_c \\
		0 \quad \quad ~ \text{for } T>T_c.
	\end{cases}
\end{equation}
In a finite-size system, the partition function $Z$ can be evaluated via the transfer matrix method, and in the leading order, $Z= \Lambda_0^L+\Lambda_1^L $, where $\Lambda_0$ and $\Lambda_1$ are the largest eigenvalue and the next-largest eigenvalue of the row transfer matrix\cite{baxter2016exactly}. In particular, for $T>T_c$, the correlation length $\xi$ is controlled by the ratio between these two eigenvalues via $\xi = \frac{1}{ \log \left( \Lambda_1/\Lambda_0  \right)  }  $. Therefore, when approaching to $T_c$ from above, the partition function is $Z= \Lambda_0^L \left( 1+ e^{-L/\xi}   \right)$, and the topological negativity behaves as $	E_{\textrm{topo}}  = \log\left( 1+ e^{-L/\xi}  \right) \quad \text{for }T>T_c.$

\subsection{Entanglement negativity when bulk excitations are allowed}\label{appendix:3d_bulk}

We here discuss the details on entanglement negativity when bulk excitations are allowed, but any point-like charges are prohibited (via $\lambda_A\to \infty$) so that the topological order persists up to a certain critical temperature. In this case, we consider the partially transposed Gibbs state $\hat{\rho}^{T_\mA}=   e^{ - \beta ( \hat{H}_{\mA } +  \hat{H}_{\mB}    )   }  \left[ e^{ -\beta \hat{H}_{\partial }}   \right]^{T_\mA}   /Z$, where $Z$ is the thermal partition function $\tr e^{ -\beta ( \hat{H}_\mA +\hat{H}_\mB +\hat{H}_{\partial})}$ and the spectrum of the partial transpose of the boundary Gibbs state $ \left[ e^{ -\beta \hat{H}_{\partial}}   \right]^{T_\mA} $ is given by Eq.\ref{eq:spectrum_A}. Here we employ a replica trick by studying $ \tr \left[\hat{\rho}^{T_{\mA}}  \right]^n$ for even integer $n$, from which negativity can be obtained by taking $n\to 1$ limit, i.e. $E_N=  \log\norm{\hat{\rho}^{T_\mA}}_1  $ with  $\norm{\hat{\rho}^{T_\mA}}_1 =  \lim_{  \text{even}~ n  \to 1 }   \tr \left[\hat{\rho}^{T_{\mA}}  \right]^n$. Specifically,  
\begin{equation}
	\begin{split}
		\norm{\hat{\rho}^{T_{\mA}}}_1 &=\lim_{\textrm{even } n  \to 1}\frac{\tr \left\{  e^{-\beta (\hat{H}_\mathcal{A}+\hat{H}_{\mathcal{B}})}   \left[  \left(e^{-\beta \hat{H}_{\partial}}\right) ^{T_{ \mathcal{A}}}\right]^n  \right\} }{  \tr   \left[ e^{-\beta \left(   \hat{H}_\mathcal{A}+\hat{H}_{\mathcal{B}}+\hat{H}_{\partial} \right)}   \right]  } \\ 
		&=\lim_{\textrm{even } n  \to 1} \frac{   \sum_{\{A_s\}} \sum_{\{ B_p \}} f(\{B_p  \}    )  e^{    -\beta(H_\mathcal{A}+H_{\mathcal{B}}) }  \cosh^{L^2}(\beta \lambda_A)     \left[     \sum_{ \{ \tau_{i} \}   }  \prod_i \tau_i^{  \frac{1-A_i }{2}}   e^{ \beta \lambda_B \sum_{\expval{ij}} B_{ij} \tau_i \tau_j  } \right]^n   }{       \sum_{\{A_s\}} \sum_{\{ B_p \}}  f(  \{B_p  \}    ) e^{    - \beta(H_\mathcal{A}+H_{\mathcal{B}}+H_{\partial}) } },
	\end{split}
\end{equation}
where the trace has been replaced by a sum over stabilizer $\{A_s\}$ and $\{B_p\}$,  subject to the local constraint that the product of  six $B_p$ stabilizers on the boundary of each cube is one: 
\begin{equation}
	f( \{B_p  \} ) = \prod_{\text{cube}} \delta\left(   \prod_{p\in \partial\text{cube} } B_p=1 \right).
\end{equation}
Also note that we have expressed the negativity spectrum of the boundary Gibbs state in terms of correlation functions of Ising spins in 2d.

By considering the limit $\lambda_A\to \infty $, every star stabilizers $A_s$ is pinned at 1 in the denominator and for the  numerator, only $A_s$ in the bulk is pinned at 1, i.e. the boundary star stabilizers $A_i$ are allowed to fluctuate. As a result,  

\begin{equation}
	\norm{\hat{\rho}^{T_{\mA}}}_1 =  \lim_{\textrm{even  }n \to 1}   2^{-L^2}    \frac{ \sum_{\{ A_i  \}} \sum^{\text{bulk}}_{\{B_p \}}  \sum_{  \{ B_{ij} \} } f(\{ B_p\})  e^{\beta \lambda_B   \sum^{\text{bulk}}_p  B_p}      \left(   \sum_{\{  \tau_i  \}}  \prod_{i}A_{i}^{\frac{1-\tau_{i}}{2}}   e^{\beta \lambda_B  \sum_{\expval{ij}} B_{ij} \tau_{i}\tau_j   }   \right)^n        }{  \sum_{\{B_p\}}  f(\{ B_p\}) e^{ \sum_{p}\beta \lambda_B  B_p}          }.
\end{equation}
Introducing $n$ copies of the Ising spins $\{ \tau_i^{\alpha}  \}$ with replica index $\alpha=1,2, \cdots, n$, one can sum over $\{A_i\}$ in the numerator: 
\begin{equation}
	\sum_{\{ A_i\}}    \left(   \sum_{\{  \tau_i  \}}  \prod_{i}A_{i}^{\frac{1-\tau_{i}}{2}}   e^{\beta \lambda_B  \sum_{\expval{ij}} B_{ij} \tau_{i}\tau_j   }   \right)^n 
	=2^{L^2} \sum_{\{\tau^{1}_i \}  }\cdots  \sum_{\{\tau^{n}_i \}  } \prod_i \delta\left(  \prod_{\alpha=1}^n\tau_i^{\alpha} =1   \right)    e^{\beta \lambda_B \sum_{\alpha=1}^n   \sum_{\expval{ij}} B_{ij}  \tau^{\alpha}_{i}\tau_j^{\alpha}   }.
\end{equation}
This is essentially a partition function for $n$ coplies of the 2d Ising model, where the spins in different replicas at any given lattice site index $i$ are coupled through the delta function constraint. Therefore, the negativity  is given by $E_N = \log \norm{ \rho^{T_{\mathcal{A}}} }_1 $ with


\begin{equation}
	\norm{ \hat{\rho}^{T_{\mathcal{A}}} }_1 =  \frac{     \sum^{\text{bulk}}_{\{B_p \}}  \sum_{  \{ B_{ij} \} } f(\{ B_p\})  e^{\beta \lambda_B   \sum^{\text{bulk}}_p  B_p}  \lim_{\textrm{even }n \to 1} \left[ \sum_{\{\tau^{1}_i \}  }\cdots  \sum_{\{\tau^{n}_i \}  }     \prod_i \delta\left(  \prod_{\alpha=1}^n\tau_i^{\alpha} =1   \right)    e^{\beta \lambda_B \sum_{\alpha=1}^n   \sum_{\expval{ij}} B_{ij}  \tau^{\alpha}_{i}\tau_j^{\alpha}   }  \right]  }{     \sum_{\{B_p\}}  f(\{ B_p\}) e^{\beta \lambda_B   \sum_p B_p}   }. 
\end{equation}

It is useful to express the result above  in terms of the ratio of two partition functions with an annealed average over the bulk fluctuations of $B_p$ stabilizers that are described by the 3d Ising gauge theory:

\begin{equation}
	\norm{ \hat{\rho}^{T_{\mathcal{A}}} }_1 =  \frac{\expval{     \sum_{  \{ B_{ij} \} }    \lim_{\textrm{even }n \to 1} \left[ \sum_{\{\tau^{1}_i \}  }\cdots  \sum_{\{\tau^{n}_i \}  }     \prod_i \delta\left(  \prod_{\alpha=1}^n\tau_i^{\alpha} =1   \right)    e^{\beta \lambda_B \sum_{\alpha=1}^n   \sum_{\expval{ij}} B_{ij}  \tau^{\alpha}_{i}\tau_j^{\alpha}   }  \right]     }_{\textrm{bulk}}      }{   \expval{   \sum_{\{ B_{ij}  \}}   e^{ \beta \lambda_B \sum_{ij   }  B_{ij} }   }_{\textrm{bulk}}    }
\end{equation}
where the numerator is 

\begin{equation}
	\begin{split}
		&\expval{     \sum_{  \{ B_{ij} \} }    \lim_{\textrm{even }n \to 1} \left[ \sum_{\{\tau^{1}_i \}  }\cdots  \sum_{\{\tau^{n}_i \}  }     \prod_i \delta\left(  \prod_{\alpha=1}^n\tau_i^{\alpha} =1   \right)    e^{\beta \lambda_B \sum_{\alpha=1}^n   \sum_{\expval{ij}} B_{ij}  \tau^{\alpha}_{i}\tau_j^{\alpha}   }  \right]     }_{\textrm{bulk}} \\
		&=  \frac{     \sum^{\text{bulk}}_{\{B_p \}}  \sum_{  \{ B_{ij} \} } f(\{ B_p\})  e^{\beta \lambda_B   \sum^{\text{bulk}}_p  B_p}  \lim_{\textrm{even }n \to 1} \left[ \sum_{\{\tau^{1}_i \}  }\cdots  \sum_{\{\tau^{n}_i \}  }     \prod_i \delta\left(  \prod_{\alpha=1}^n\tau_i^{\alpha} =1   \right)    e^{\beta \lambda_B \sum_{\alpha=1}^n   \sum_{\expval{ij}} B_{ij}  \tau^{\alpha}_{i}\tau_j^{\alpha}   }  \right]  }{     \sum^{\text{bulk}}_{\{B_p \}}   f(\{ B_p\})  e^{\beta \lambda_B   \sum^{\text{bulk}}_p  B_p}  }. 
	\end{split}
\end{equation}
and the denominator is

\begin{equation}
	\expval{   \sum_{\{ B_{ij}  \}}   e^{ \beta \lambda_B \sum_{ij   }  B_{ij} }   }_{\textrm{bulk}}    =   \frac{ \sum^{\text{bulk}}_{\{B_p \}}    f(\{ B_p\})  e^{\beta \lambda_B   \sum^{\text{bulk}}_p  B_p}  \left[ \sum_{\{ B_{ij}  \}}   e^{ \beta \lambda_B \sum_{ij   }  B_{ij} }      \right] }{   \sum^{\text{bulk}}_{\{B_p \}}   f(\{ B_p\})  e^{\beta \lambda_B   \sum^{\text{bulk}}_p  B_p}    }
\end{equation}

As a sanity check for the replica trick, one can show that when forbidding the bulk excitations, i.e. $B_p=1$ in the bulk, $\norm{ \hat{\rho}^{T_{\mathcal{A}}} }_1 $ recovers the result in  Appendix.\ref{appendix:3d_exact}. In this case,

\begin{equation}
	\norm{ \hat{\rho}^{T_{\mathcal{A}}} }_1 =  \frac{ \sum_{  \{ B_{ij} \} }    \lim_{\textrm{even }n \to 1} \left[ \sum_{\{\tau^{1}_i \}  }\cdots  \sum_{\{\tau^{n}_i \}  }     \prod_i \delta\left(  \prod_{\alpha=1}^n\tau_i^{\alpha} =1   \right)    e^{\beta \lambda_B \sum_{\alpha=1}^n   \sum_{\expval{ij}} B_{ij}  \tau^{\alpha}_{i}\tau_j^{\alpha}   }  \right]         }{     \sum_{\{ B_{ij}  \}}   e^{ \beta \lambda_B \sum_{ij   }  B_{ij} }    }
\end{equation}
with $\{B_{ij}\}$ subject to the constraint that $\prod_{\expval{ij}   \in \partial p    }  B_{ij }=1   $ on the 2d bipartition boundary. Using the local gauge symmetry in the numerator: $B_{ij} \to - B_{ij}$ for four links emanating from a site $i$ with  $\tau_i^{\alpha} \to  -\tau_i^{\alpha}   $ for all replicas $\alpha=1, 2, \cdots, n$, one can remove $B_{ij}$ in the Gibbs weight, the numerator  can be simplified as

\begin{equation} 
	\sum_{  \{ B_{ij} \} }    \lim_{\textrm{even }n \to 1} \left[ \sum_{\{\tau^{1}_i \}  }\cdots  \sum_{\{\tau^{n}_i \}  }     \prod_i \delta\left(  \prod_{\alpha=1}^n\tau_i^{\alpha} =1   \right)    e^{\beta \lambda_B \sum_{\alpha=1}^n   \sum_{\expval{ij}}  \tau^{\alpha}_{i}\tau_j^{\alpha}   }  \right]     =     \sum_{  \{ B_{ij} \} }  \sum_{\{\tau_i \}  }     \delta\left( \tau_i=1  \right) e^{\beta \lambda_B   \sum_{\expval{ij}}  \tau_{i}\tau_j }  = \sum_{  \{ B_{ij}  \} }  e^{ \beta \lambda_B 2L^2}. 
\end{equation} 
Note that such the aforementioned gauge symmetry exists only for even $n$ while for odd $n$, sending $\tau_i^{\alpha} $ to $  -\tau_i^{\alpha}   $ for all replicas is not allowed (due to the violation of the constraint $\prod_{\alpha=1}^n \tau_i^{\alpha}=1$). On the other hand, natively taking $n\to 1$ would lead to $ \sum_{\{ B_{ij}  \}}   e^{ \beta \lambda_B \sum_{ij   }  B_{ij} } $ for the numerator, which would give $\norm{ \hat{\rho}^{T_\mA }  }_1=1$ (i.e. negativity would be zero).

Using the constraint $\prod_{\expval{ij}   \in \partial p    }  B_{ij }=1   $, one can   introduce the Ising variables via $B_{ij} = \tau_i \tau_j $ so that 

\begin{equation}
	\norm{\hat{\rho}^{T_\mA}}_1 = \frac{ \sum_{\{  \tau_{ i }    \} }  e^{2L^2 \beta 
			\lambda_B}    }{    \sum_{\{ \tau_l\} }  e^{ \beta    \lambda_B  \sum_{\expval{ij} }\tau_i\tau_j   }    }  = \frac{2^{L^2}  e^{ 2\beta \lambda_B L^2}    }{Z}
\end{equation}
with $Z=   \sum_{\{ \tau_l\} }  e^{ \beta    \lambda_B  \sum_{\expval{ij} \tau_i\tau_j  }  }  $ being the partition function of 2d Ising model that behaves as $Z= 2 e^{ -\beta L^2 f }$ for $T<T_c$ and  $e^{ -\beta L^2 f }$ for $T>T_c$ with $f$ being the free energy density. Therefore, the non-zero $\log 2$ topological entanglement negativity results from the spontaneous symmetry breaking of the Ising model that emerges from the local constraint of the boundary $B_{ij}$ stabilizers, namely, $\prod_{\expval{ij}   \in \partial p    }  B_{ij }=1 $. 

Now  we consider the Gibbs state with tunable temperature in the bulk and the boundary, i.e. $\hat{\rho}\sim e^{ -\beta_{\text{bulk}} (H_{\mA} + H_{\mB}  )   } e^{ - \beta_{\partial} H_{\mA \mB}}$. The 2d boundary theory is coupled to the 3d Ising gauge theory in the bulk, and the aforementioned constraint of the boundary plaquettes no longer exists due to the fluctuating $B_{p}$ stabilizers in the bulk. However, when the bulk is in the low-temperature deconfined phase ($T_{\textrm{bulk}}< T_{\textrm{bulk},c}  $), the Wilson loop operator satisfies the perimeter law, i.e. $\expval{W(C)} \sim e^{ -\alpha |C| }$ where $|C|$ is the perimeter of the closed loop $C$. It is known that one can find a renormalized (fattened) Wilson loop operator such that it satisfies the zero-law (i.e. the Wilson loop does not decay at all). Since the product of plaquette across the bipartition boundary is equivalent to the product of two Wilson loops in the bulk, one expects to find an emergent constraint on those boundary plaquettes satisfied on a larger length scale so that one can find a coarse-grained 2d Ising model, which displays an order-disorder transition as tuning the boundary temperature. On the other hand, for $T_{\textrm{bulk}} > T_{\textrm{bulk},c}$, due to the confinement of the Wilson loop in the bulk, the emergent constraint on the boundary plaquettes no longer exists. Therefore, there is no emergent 2d Ising model description that exhibits an ordered phase, contributing to topological entanglement negativity.

\section{Structure of negativity spectrum and  entanglement negativity  in 4d toric code}\label{appendix:4d}

\subsection{Structure of negativity spectrum }

The 4d toric code exhibits a topological order below a certain critical temperature $T_c$. As a simplification, here we consider only the bipartition-boundary part of the density matrix by forbidding any excitations in the bulk. The spectrum of the partially transposed boundary part of the Gibbs state is characterized by strange correlators that can be written as correlation functions in a 3d Ising gauge theory coupled to matter field: 
\begin{equation}\label{appendix:4d_spectrum}
	\left[  e^{-\beta H_{\partial}} \right]^{T_\mA} \sim  \bra{+} \left[ \prod_{l} Z_l^{ \frac{1-A_l}{2}}   \right]  \left[ \prod_{p} Z_p^{ \frac{1-B_p}{2}}   \right] \ket{\psi(T)} \sim   \sum_{   \{ \tau_{l} \} }  \left[\prod_l   {\tau_l}^{\frac{1-A_l   }{2}  }  \right] e^{ -K_A \sum_l \frac{1-\tau_l}{2}  + \beta \lambda_B  \sum_p B_p \prod_{l\in \partial p} \tau_l   }
\end{equation}
with $K_A = - \log[  \tanh(\beta \lambda_A)   ]$. Here $\{A_l\}$ determines the spin insertion in the correlator and  $\{B_{p} \}$ determines the sign of interaction between spins in the 3d lattice. To understand the structure of negativity spectrum, it is useful to denote $A_l=-1$ with  an occupied link in the lattice and denote $B_p=-1$ with an occupied link in the dual lattice piercing through the plaquette $p$ on the original lattice. Due to the local constraints in 4d toric code, the bipartition-boundary stabilizers $A_l$ and $B_p$ are subject to the constraints that the product of 6 plaquettes $B_p$ on the boundary of each cube is one, and the product of 6 links $A_l$ emanating from each vertex is one. The above constraints amount to imposing the condition that only closed loops of $A_l$ (denoted as $A$-loops) in the direct lattice and closed loops of $B_p$ (denoted as $B$-loops) in the dual lattice are allowed. Eq.\ref{appendix:4d_spectrum} shows that the negativity spectrum is characterized by a classical 3d Ising gauge theory coupled to matter fields, which therefore exhibits a deconfinement-confinement transition at a certain critical temperature. As a result, such a transition corresponds to the transition for topological order in the toric code. Note that it is interesting that while the thermal partition function $\tr e^{ -\beta \hat{H}_{\mA \mB} }$ can be written as a product of two partition functions for two independent pure gauge theories (therefore exhibiting two transitions with critical temperatures $T_A\sim O(\lambda_A)$ and $T_B \sim O(\lambda_B)$), the partially transposed Gibbs state exhibits a single deconfined transition which is determined by both $\lambda_A$ and $\lambda_B$. It is also interesting that setting $\lambda_A=\lambda_B$ while vaying the temperature corresponds to a transition along the well-known self-dual line in the gauge theory\cite{fradkin_shenker_diagram,gauge_mc_calculation_Jayaprakash,gauge_phase_diagram_Stamp,gauge_phase_diagram_vidal,nahum_dual_2020}

\subsection{Exact  entanglement negativity}\label{sec:4d_exact}
Here we consider the limit $\beta \lambda_A\to \infty$, i.e. $K_A=0$, and discuss the sign structure of negativity spectrum and the derivation of entanglement negativity. In this case, the negativity spectrum is given by the pure gauge theory 

\begin{equation}
	\left[  e^{-\beta H_{ \partial}} \right]^{T_\mA} \sim   \sum_{   \{ \tau_{l} \} }  \left[\prod_l   {\tau_l}^{\frac{1-A_l   }{2}  }  \right]   e^{ \beta \lambda_B  \sum_p B_p \prod_{l\in \partial p} \tau_l   }.
\end{equation}

Here  we first discuss the braiding structure between $A$-loops and $B$-loops. Consider the case with a single $A$-loop, the corresponding eigenvalue is $ \sum_{   \{ \tau_{l} \} } \left[\prod_{ l \in C_A  } \tau_l \right] e^{ \beta \lambda_B  \sum_p  \prod_{l\in \partial p} \tau_l   }$, which is nothing but a Wilson loop in the 3d Ising gauge theory. Such a quantity exhibits a long-range correlation below a certain critical temperature $T_c$ (deconfined phase) in the sense of a perimeter-law $e^{  -  \alpha  |C_A| }  >0 $, where $|C_A|$ denotes the length of the close loop $C_A$. Now we consider adding a $B$-loop in the dual lattice that braids with the $A$-loop by flipping $B_p$. It follows that the corresponding eigenvalue can be written as $ -\sum_{   \{ \tau_{l} \} } \left[\prod_{ l \in C_A  } \tau_l \right] e^{  \beta \lambda_B  \sum_p  \prod_{l\in \partial p} \tau_l   }$, which  remains long-range correlated in the deconfined phase with the perimeter-law scaling $-e^{  -  \alpha  |C_A| }  >0 $. The above analysis indicates that the braiding and sign structure survives in the long-distance below $T_c$.

Now we discuss the calculation of negativity. First, the one norm of $\hat{\rho}^{T_\mA}$ with $\hat{\rho} \sim e^{ -\beta \hat{H}_{\partial}   }$ is the sum of all absolute eigenvalues:

\begin{equation}
	\norm{\hat{\rho}^{T_\mA}}_1= \frac{ \sum'_{\{A_l   \},\{ B_p \} }  \abs{ \sum_{  \{\tau_l   \}  }\prod_{ l}   A_l^{\frac{1-\tau_l}{2}}   e^{ \beta \lambda_B \sum_p  B_p \prod_{ l \in \partial p } \tau_l }} }{ \sum'_{\{A_l   \},\{ B_p \} } \sum_{  \{\tau_l   \}  }\prod_{ l} A_l^{\frac{1-\tau_l}{2}}   e^{ \beta \lambda_B \sum_p  B_p \prod_{l \in \partial p }  \tau_l      }  },
\end{equation}
where the denominator is simply fixed by the requirement that sum of all eigenvalues of $\rho^{T_\mA}$ is one. Here $\sum'_{\{A_l\}   \{B_{p}\}}$ refers to summing over $\{A_l\}$ and  $\{B_{p } \}  $ subject to the constraints that the product of 6 plaquettes $B_p$ on the boundary of each cube is one, and the product of 6 links $A_l$ emanating from each vertex is one. To resolve the local constraint of $\{B_p\}$, one introduces the dual variables $\{g_l\}$ on links via $B_p= \prod_{l\in \partial p } g_l$, and therefore $\sum_{\{B_p\}  } $ can be replaced by summing over independent $g_l$ variables. Using a calculation analogous to 3d toric code, we find

\begin{equation}
	\norm{\hat{\rho}^{T_\mA}}_1 = \frac{ \sum_{\{  g_{ l }    \} }  e^{3N \beta 
			\lambda_B}    }{    \sum_{\{ g_l\} }  e^{ \beta    \lambda_B  \sum_p  \prod_{l  \in \partial p } g_{ l }  }    }  =  \frac{  2^{3N} e^{- \beta E_g} }{Z(T)}
\end{equation}
where $N=L^3$, $E_g = - 3N \lambda_B$ and  $Z(T)=   \sum_{\{ \tau_l\} }  e^{ \beta    \lambda_B  \sum_p  \prod_{l  \in \partial p }  g_l  }    $ is ground state energy and the partition function for the 3d pure $\mathbb{Z}_{2}$ gauge theory. Taking a logarithm gives the entanglement negativity

\begin{equation}
	E_N= 3N\log 2 - \beta E_g  -\log Z(T). 
\end{equation}
This expression allows us to compute the topological negativity $E_{\textrm{topo}}$, i.e.   the  long-range component of entanglement negativity: $E_{\textrm{topo}}=  E_N(2N)-2E_N(N)=   -\log Z(2N,T)+ 2 \log Z(N,T)$. To compute such a quantity, a simple way is by mapping the finite-temperature 3d classical Ising gauge theory to the zero-temperature 2+1D quantum Ising gauge theory with Gauss law imposed:

\begin{equation}
	Z=  \sum_{\{ g_l\} }  e^{ \beta   \sum_p  \prod_{l  \in \partial p }  g_{ l }  }   \sim  \lim_{M \to \infty}   \tr \left(  e^{  M \beta \sum_p \prod_{l \in \partial p}  Z_l +M \beta   g \sum_l X_l  }  \right) 
\end{equation}
with $e^{ -2 \beta g } =  \tanh(\beta)$. Therefore, $Z\sim N_g  e^{ -M \beta E_g    }$, where $E_g$ is the ground state energy of the 2+1D quantum Ising gauge  theory, and $N_g$ is the corresponding ground state degeneracy. Crucially, tuning $g$ in such a model induces a confinement-deconfinement transition at a critical $g_c$. The regime $g<g_c$ corresponds to the deconfined phase with $N_g=4$, while the regime $g>g_c$ corresponds to the confined phase with $N_g=1$. As a result, there exists a universal subleading term in $\log Z$ that  characterizes the number of topological sector in the gauge theory, and $E_{\text{topo}}=2\log 2$ or $0$ for $T<T_c$ and $T>T_c$.

\subsection{Replica calculation for general $\lambda_A$ and $\lambda_B$ when forbidding bulk excitations }
In the discussion above,  we consider the limit $\lambda_A\to \infty$ to derive the entanglement negativity. For the  negativity at any $\lambda_A$ and $\lambda_B$, we employ a replica trick by studying $ \tr \left[\hat{\rho}^{T_{\mA}}  \right]^n$ for even integer $n$, from which negativity can be obtained by taking $n\to 1$ limit, namely, $E_ N =  \lim_{  \text{even}~ n  \to 1 }   \tr \left[  \hat{\rho}^{T_{\mA}}  \right]^n     =  \lim_{  \text{even}~ n  \to 1 } \tilde{Z}_n/Z    $, where $Z$ is the thermal partition function $Z= \tr e^{ -\beta  \hat{H}_{\partial}  }$ and $\tilde{Z}_n$ is the n-th moment for the boudary part of the Gibbs state, i.e. $\tilde{Z}_n = \tr \left\{  	\left[  e^{-\beta  \hat{H}_{ \partial}} \right]^{T_\mA} \right\}^n$. Using the negativity spectrum (Eq.\ref{appendix:4d_spectrum}), one finds 

\begin{equation}
	\tilde{Z}_n  \sim  \sum'_{\{A_l   \},\{ B_p \} }  \left[ \sum_{  \{\tau_l   \}  }\prod_{ l} \tau_l  ^{\frac{1-A_l}{2}}   e^{ -K_A \sum_l \frac{1-\tau_l}{2}  +    \beta \lambda_B \sum_p  B_p \prod_{ l \in \partial p } \tau_l }\right]^n, 
\end{equation}
where $K_A =-\log[  \tanh( \beta \lambda_A)  ]$ and   $\sim $ indicates that we have omitted a prefactor $[\cosh(\beta  \lambda_A  )]^{N_l}$ with $N_l$ being the number of links in the 3d lattice. Due to the local constraint for $B_p$ that the product of six $B_p$ on the boundary of a cube is one, any allowed $\{B_p\}$ can be reached by flipping four $B_p$ sharinge a link $l$, which is equivalently to flipping the spin $\tau_l$. Therefore, one can introduce independent $g_l =\pm 1$ variables living on links to resolve the local constraint on $\{B_p\}$ and find

\begin{equation}
	\tilde{Z}_n  \sim  \sum'_{\{A_l   \},\{  g_l \} }  \left[ \sum_{  \{\tau_l   \}  }\prod_{ l}  \tau _l ^{\frac{1-A_l }{2}}   e^{ -K_A \sum_l \frac{1- g_l \tau_l}{2}  +    \beta \lambda_B \sum_p  \prod_{ l \in \partial p } \tau_l }\right]^n.
\end{equation}
By introducing  $n$ replicas:
\begin{equation}
	\tilde{Z}_n \sim  	 \sum'_{\{A_l   \},\{g_l\} } \sum_{  \{\tau^{\alpha}_l   \}  } \prod_{ l} \left(  A_l  \right)^{ \sum_{\alpha=1}^n  \frac{1-\tau^{\alpha}_l}{2}}   e^{ -K_A \sum_{l, \alpha }  \frac{1-g_l \tau^{\alpha}_l}{2}  +    \beta \lambda_B \sum_p  \sum_{\alpha}  \prod_{ l \in \partial p } \tau^{\alpha}_l },
\end{equation}
with $\alpha$ denoting the replica index, we can sum over $\{A_l \}$ subject to the constraint that the product of six $A_l$ on links emanating from a vertex is one. This effectively couples spins on different replicas: $\sum'_{\{A_l\}}  \prod_{ l} \left(  A_l  \right)^{ \sum_{\alpha=1}^n  \frac{1-\tau^{\alpha}_l}{2}}   \sim \prod_p \delta\left( \prod_{\alpha, l \in \partial p } \tau_l^\alpha  =1  \right) $, where the constraint is that for any given plaquette $p$, the product of spins on its boundary across all replicas is one. Therefore, one finally simplifies $\tilde{Z}_n = \tr \left\{  	\left[  e^{-\beta H_{\partial}} \right]^{T_\mA} \right\}^n$ as 

\begin{equation}
	\tr \left\{  	\left[  e^{-\beta H_{\partial}} \right]^{T_\mA} \right\}^n \sim  \sum_{\{g_l\}}  \sum_{  \{\tau^{\nu}_l   \}  }  e^{ -K_A \sum_{l, \nu }  \frac{1-g_l \tau^{\nu}_l}{2}  +    \beta \lambda_B \sum_p  \sum_{\nu}  \prod_{ l \in \partial p } \tau^{\nu}_l } \prod_p \delta\left( \prod_{\nu, l \in \partial p } \tau_l^\nu  =1  \right) ,
\end{equation}
and evaluating such a quantity for even $n$ followed an analytic continuation to $n\to 1$  gives entanglement negativity. Although we are unable to compute such an quantity analytically, this expression suggests that the transition in  negativity relates to the deconfinement-confinement transition of 3d Ising gauge theory coupled to matter fields.

\subsection{Entanglement negativity when bulk is thermal}

We here discuss the entanglement negativity when bulk excitations are allowed, but one type of excitations is prohibited (via $\lambda_A\to \infty$). In this case, we consider the partially transposed Gibbs state $\hat{\rho}^{T_\mA}=   e^{ - \beta ( \hat{H}_{\mA } +  \hat{H}_{\mB}    )   }  \left[ e^{ -\beta \hat{H}_{\partial }}   \right]^{T_\mA}   /Z$, where $Z$ is the thermal partition function $\tr e^{ -\beta ( \hat{H}_\mA +\hat{H}_\mB +\hat{H}_{\partial})}$ and the spectrum of the partial transpose of the boundary Gibbs state $ \left[ e^{ -\beta \hat{H}_{\partial }}   \right]^{T_\mA} $ is given by Eq.\ref{appendix:4d_spectrum}. Here we employ a replica trick to compute the negativity, i.e. $E_N=  \log\norm{\hat{\rho}^{T_\mA}}_1  $ with  $\norm{\hat{\rho}^{T_\mA}}_1 =  \lim_{  \text{even}~ n  \to 1 }   \tr \left[\hat{\rho}^{T_{\mA}}  \right]^n$. Specifically,  
\begin{equation}
	\begin{split}
		\norm{\hat{\rho}^{T_{\mA}}}_1 &=\lim_{\textrm{even } n  \to 1}\frac{\tr \left\{  e^{-\beta (\hat{H}_\mathcal{A}+\hat{H}_{\mathcal{B}})}   \left[  \left(e^{-\beta \hat{H}_{\partial}}\right) ^{T_{ \mathcal{A}}}\right]^n  \right\} }{  \tr   \left[ e^{-\beta \left(   \hat{H}_\mathcal{A}+\hat{H}_{\mathcal{B}}+\hat{H}_{\partial} \right)}   \right]  } \\ 
		&=\lim_{\textrm{even } n  \to 1} \frac{   \sum_{\{A_l\}} \sum_{\{ B_c \}} f( \{  A_l\},\{B_c \}    )  e^{    -\beta(H_\mathcal{A}+H_{\mathcal{B}}) }  \cosh^{3L^3}(\beta \lambda_A)     \left[     \sum_{ \{ \tau_l \}   }  \prod_l \tau_l^{  \frac{1-A_l }{2}}   e^{ \beta \lambda_B \sum_{ p   } B_{p} \prod_{l\in \partial p} \tau_l  } \right]^n   }{       \sum_{\{A_s\}} \sum_{\{ B_p \}}   f( \{  A_l\},\{B_p  \}    ) e^{    - \beta(H_\mathcal{A}+H_{\mathcal{B}}+H_{ \partial}) } },
	\end{split}
\end{equation}
where the trace has been replaced by a sum over stabilizer $\{A_l\}$ defined on 1-cells and $\{B_c\}$ defined on 3-cells in the 4d lattice,  subject to the local constraint that the product of  six $B_p$ stabilizers on the boundary of each cube is one: 
\begin{equation}
	f( \{  A_l\},\{B_c  \}    ) =  \prod_{4-\textrm{cell}} \delta\left(  \prod_{c \in \partial 4-\textrm{cell}}  B_c=1   \right)  \prod_{0-\textrm{cell}} \delta\left(   \prod_{ 0-\textrm{cell } \in \partial l   }  A_l   =1 \right).
\end{equation}
In addition, we use $B_p$ to denote the cube stabilizers $B_c$ across the 3d bipartition boundary since those stabilizers live on plaquettes of in the 3d lattice.

By considering the limit $\lambda_A\to \infty $, every stabilizer $A_l$ is pinned at 1 in the denominator and for the  numerator, only $A_l$ in the bulk is pinned at 1, i.e. the stabilizers $A_l$ across the 3d bipartition boundary fluctuate. Note that these boundary $A_l$ stabilizers are subject to the constraint that the product of six $A_l$ emanating from a vertex is one in the 3d bipartition boundary, i.e. $\prod_v \delta\left( \prod_{   v\in \partial l} A_l =1\right)$. As a result,  

\begin{equation}
	\norm{\hat{\rho}^{T_{\mA}}}_1 =  \lim_{\textrm{even  }n \to 1}   2^{-3L^3}    \frac{  \sum^{\text{bulk}}_{\{B_c \}}    \sum^{\partial}_{\{ A_l  \}}  \sum_{  \{ B_{p} \} } f( \{  A_l\},\{B_c  \}    )  e^{\beta \lambda_B   \sum^{\text{bulk}}_c B_c}      \left(   \sum_{\{  \tau_l \}}  \prod_lA_l^{\frac{1-\tau_l}{2}}   e^{\beta \lambda_B  \sum_{p} B_p  \prod_{l\in \partial p} \tau_l   }   \right)^n        }{  \sum_{\{B_c\}}  f(\{ B_c\}) e^{ \sum_{c}\beta \lambda_B  B_c}          }.
\end{equation}
Introducing $n$ copies of the Ising spins $\{ \tau_l^{\alpha}  \}$ with $\alpha=1,2, \cdots, n$, one can sum over $\{A_l\}$ in the numerator, subject to the constraint  $\prod_v \delta\left( \prod_{   v\in \partial l} A_l =1\right)$: 
\begin{equation}
	\sum_{\{ A_l\}}    \left(   \sum_{\{  \tau_l  \}}  \prod_{l}A_{l}^{\frac{1-\tau_l}{2}}   e^{\beta \lambda_B  \sum_{p} B_p \prod_{ l \in \partial p} \tau_l   }   \right)^n 
	=2^{3L^3} \sum_{\{\tau^{1}_l \}  }\cdots  \sum_{\{\tau^{n}_l \}  }  
	\prod_p \delta\left( \prod_{\alpha, l \in \partial p } \tau_l^\alpha  =1  \right)	   e^{\beta \lambda_B \sum_{\alpha=1}^n   \sum_{p} B_p  \prod_{l \in \partial p } \tau_l^{\alpha}    }   ,
\end{equation}
where different replicas are coupled by the constraint that  for any given plaquette $p$, the product of spins its boundary across all replicas is one. Therefore, the negativity  is given by $E_N = \log \norm{ \hat{\rho}^{T_{\mathcal{A}}} }_1 $ with 

\begin{equation}
	\norm{ \hat{\rho}^{T_{\mathcal{A}}} }_1 =  \frac{     \sum^{\text{bulk}}_{\{B_c\}}  \sum_{ \{B_p\} } f(\{ B_p\})  e^{\beta \lambda_B   \sum^{\text{bulk}}_c  B_c}  \lim_{\textrm{even }n \to 1} \left[\sum_{\{\tau^{1}_l \}  }\cdots  \sum_{\{\tau^{n}_l \}  }  
		\prod_p \delta\left( \prod_{\alpha, l \in \partial p } \tau_l^\alpha  =1  \right)	   e^{\beta \lambda_B \sum_{\alpha=1}^n   \sum_{p} B_p  \prod_{l \in \partial p } \tau_l^{\alpha}    }  \right]  
	}{     \sum_{\{B_c\}}  f(\{ B_c\}) e^{\beta \lambda_B   \sum_c B_c}   }. 
\end{equation}

It is useful to express the result above  in terms of the ratio of two partition functions with an annealed average over the bulk fluctuations of $B_c$ stabilizers that are described by the 4d Ising gauge theory:

\begin{equation}
	\norm{ \hat{\rho}^{T_{\mathcal{A}}} }_1 =  \frac{     \expval{     \sum_{  \{B_p \} }    \lim_{\textrm{even }n \to 1} \left[ \sum_{\{\tau^{1}_l \}  }\cdots  \sum_{\{\tau^{n}_l \}  }  
			\prod_p \delta\left( \prod_{\alpha, l \in \partial p } \tau_l^\alpha  =1  \right)	   e^{\beta \lambda_B \sum_{\alpha=1}^n   \sum_{p} B_p  \prod_{l \in \partial p } \tau_l^{\alpha}    }   \right]     }_{\textrm{bulk}}      }{   \expval{   \sum_{\{ B_{p}  \}}   e^{ \beta \lambda_B \sum_{p   }  Bp }   }_{\textrm{bulk}}    }
\end{equation}
where the numerator is 

\begin{equation}
	\begin{split}
		&  \expval{     \sum_{  \{B_p \} }    \lim_{\textrm{even }n \to 1} \left[ \sum_{\{\tau^{1}_l \}  }\cdots  \sum_{\{\tau^{n}_l \}  }  
			\prod_p \delta\left( \prod_{\alpha, l \in \partial p } \tau_l^\alpha  =1  \right)	   e^{\beta \lambda_B \sum_{\alpha=1}^n   \sum_{p} B_p  \prod_{l \in \partial p } \tau_l^{\alpha}    }   \right]     }_{\textrm{bulk}}  \\
		&=  \frac{    \sum^{\text{bulk}}_{\{B_c\}}  \sum_{ \{B_p\} } f(\{ B_p\})  e^{\beta \lambda_B   \sum^{\text{bulk}}_c  B_c}  \lim_{\textrm{even }n \to 1} \left[\sum_{\{\tau^{1}_l \}  }\cdots  \sum_{\{\tau^{n}_l \}  }  
			\prod_p \delta\left( \prod_{\alpha, l \in \partial p } \tau_l^\alpha  =1  \right)	   e^{\beta \lambda_B \sum_{\alpha=1}^n   \sum_{p} B_p  \prod_{l \in \partial p } \tau_l^{\alpha}    }  \right]   }{     \sum^{\text{bulk}}_{\{B_c \}}   f(\{ B_c\})  e^{\beta \lambda_B   \sum^{\text{bulk}}_c  B_c}  }. 
	\end{split}
\end{equation}
and the denominator is

\begin{equation}
	\expval{   \sum_{\{ B_p  \}}   e^{ \beta \lambda_B \sum_p  B_p }   }_{\textrm{bulk}}    =   \frac{ \sum^{\text{bulk}}_{\{B_c \}}    f(\{ B_c\})  e^{\beta \lambda_B   \sum^{\text{bulk}}_c  B_c}  \left[ \sum_{\{ B_p  \}}   e^{ \beta \lambda_B \sum_p B_p }      \right] }{   \sum^{\text{bulk}}_{\{B_c \}}   f(\{ B_c\})  e^{\beta \lambda_B   \sum^{\text{bulk}}_c  B_c}    }
\end{equation}

As a sanity check for the replica trick, one can show that when forbidding the bulk excitations, i.e. $B_c=1$ in the bulk, $\norm{ \hat{\rho}^{T_{\mathcal{A}}} }_1 $ recovers the result in  Appendix. \ref{sec:4d_exact}. In this case,

\begin{equation}
	\norm{ \hat{\rho}^{T_{\mathcal{A}}} }_1 =  \frac{ \sum_{  \{ B_p \} }      \lim_{\textrm{even }n \to 1} \left[ \sum_{\{\tau^{1}_l \}  }\cdots  \sum_{\{\tau^{n}_l \}  }  \prod_p \delta\left( \prod_{\alpha, l \in \partial p } \tau_l^\alpha  =1  \right)	   e^{\beta \lambda_B \sum_{\alpha=1}^n   \sum_{p} B_p  \prod_{l \in \partial p } \tau_l^{\alpha}    }   \right]         }{     \sum_{\{ B_{ij}  \}}   e^{ \beta \lambda_B \sum_p  B_p }    }
\end{equation}
with $\{B_p\}$ subject to the constraint the product of 6 $B_p$ on the boundary of each 3-cell is one, i.e. $\prod_{p   \in \partial c    }  B_p =1   $ in the 3d bipartition boundary. Using the local gauge symmetry in the numerator: $B_p \to - B_p$ for four plaquettes sharing the same boundary edge labeled by $l$ with  $\tau_l^{\alpha} \to  -\tau_l^{\alpha}  $ for all replicas $\alpha=1, 2, \cdots, n$, one can remove $B_p$ in the Gibbs weight, the numerator  can be simplified as

\begin{equation} 
	\begin{split}
		\sum_{  \{ B_p \} }    \lim_{\textrm{even }n \to 1} \left[ \sum_{\{\tau^{1}_l \}  }\cdots  \sum_{\{\tau^{n}_l \}  }     \prod_p \delta\left(  \prod_{\alpha=1}^n   \prod_{ l \in \partial p}\tau_l^{\alpha} =1   \right)    e^{\beta \lambda_B \sum_{\alpha=1}^n   \sum_p  \prod_{ l \in \partial p} \tau_l^{\alpha}  }  \right]   &  =     \sum_{  \{ B_p \} }  \sum_{\{\tau_l \}  }     \delta\left( \prod_{l \in \partial  p }\tau_l=1  \right) e^{\beta \lambda_B   \sum_{p }  \prod_{l\in \partial p } \tau_l }  \\
		& = \sum_{  \{B_p  \} }  e^{ \beta \lambda_B 3L^3}. 
	\end{split}
\end{equation} 
Note that such the aforementioned gauge symmetry exists only for even $n$ while for odd $n$, sending $\tau_i^{\alpha} $ to $  -\tau_i^{\alpha}   $ for all replicas is not allowed (due to the violation of the constraint that couples different replicas). On the other hand, natively taking $n\to 1$ would lead to $ \sum_{\{ B_p  \}}   e^{ \beta \lambda_B \sum_{p }  B_p } $ for the numerator, which would give $\norm{ \hat{\rho}^{T_\mA }  }_1=1$ (i.e. negativity would be zero).

To resolve the constraint  $\prod_{p\in \partial c} B_p=1$, one can introduce the  Ising variables $\tau_l $ defined on links via $B_{p} = \prod_{l \in \partial p} \tau_l$ so that 

\begin{equation}
	\norm{\hat{\rho}^{T_\mA}}_1 = \frac{ \sum_{\{  \tau_{ l }    \} }  e^{3L^3 \beta 
			\lambda_B}    }{    \sum_{\{ \tau_l\} }  e^{ \beta    \lambda_B  \sum_p \prod_{l \in  \partial p} \tau_l  }    }  = \frac{3^{L^3}  e^{ 3\beta \lambda_B L^3}    }{Z}
\end{equation}
with $Z=   \sum_{\{ \tau_l\} }  e^{ \beta    \lambda_B  \sum_p \prod_{l \in  \partial p} \tau_l  }  $ being the partition function of 3d Ising pure gauge theory. By mapping this partition function at finite temperature to the zero-temperature 2+1D quantum Ising gauge theory, one finds $\log Z$ has a universal subleading term $2\log 2, 0  $ for $T<T_c$ and $T>T_c$. Such a term  corresponds to the topological entanglement negativity results from the deconfinement  transition  of the Ising gauge theory that emerges from the local constraint of the boundary $B_p$ stabilizers, namely, $\prod_{ p \in \partial c}  B_p=1 $. 

Now  we consider the Gibbs state with tunable temperature in the bulk and the boundary, i.e. $\hat{\rho}\sim e^{ -\beta_{\text{bulk}} (\hat{H}_{\mA} + \hat{H}_{\mB}  )   } e^{ - \beta_{\partial} \hat{H}_{\partial}}$. The 3d boundary theory is coupled to the 4d Ising gauge theory in the bulk, and the aforementioned constraint of the boundary plaquettes no longer exists due to the fluctuating $B_{c}$ stabilizers in the bulk. However, when the bulk is in the low-temperature deconfined phase ($T_{\textrm{bulk}}< T_{\textrm{bulk},c}  $), one can still find an emergent constraint on the $B_p$ stabilizers in the 3d bipartition boundary (similar to the discussion for 3d toric code in Appendix.\ref{appendix:3d_bulk}) to derive a coarse-grained 3d Ising gauge theory that exhibits a deconfinement-confinement transition as tuning the boundary temperature $T_{\partial}$. This imlpies that for $T_{\textrm{bulk}}< T_{\textrm{bulk},c}$, tuning the boundary temperature  $T_{\partial}$ still drives a transition from long-range entangled phase to a short-range entangled phase whose universality is the same as the case with $T_{\textrm{bulk}} = 0 $.

\end{document}